\documentclass[10pt]{article}
\usepackage[utf8]{inputenc}
\usepackage[affil-it]{authblk}
\usepackage[normalem]{ulem}
\usepackage{bm}
\usepackage{enumitem}
\usepackage{amsmath}
\usepackage{graphicx}
\usepackage{geometry}
\usepackage{amssymb}
\usepackage{hyperref,cite}
\usepackage{subcaption}
\usepackage{placeins}
\usepackage{mathtools}
\usepackage{mathrsfs}
\usepackage{csquotes}
\usepackage{comment}
\usepackage{multirow}
\usepackage{array,soul}
\usepackage[dvipsnames]{xcolor}
\providecommand{\keywords}[1]{\textbf{\textit{Keywords---}} #1}

\title{Multiscale Nonlocal Elasticity:\\ A Distributed Order Fractional Formulation}
\author{Wei Ding\thanks{To whom correspondence should be addressed. Email: ding242@purdue.edu or fsemperl@purdue.edu}}
\author{Sansit Patnaik}
\author{Fabio Semperlotti${}^*$}
\affil{Ray W. Herrick Laboratories, School of Mechanical Engineering, Purdue University, West Lafayette, IN 47907}

\begin{document}
\date{}
\maketitle
\begin{abstract}
This study presents a generalized multiscale nonlocal elasticity theory that leverages distributed order fractional calculus to accurately capture coexisting multiscale and nonlocal effects within a macroscopic continuum. The nonlocal multiscale behavior is captured via distributed order fractional constitutive relations derived from a nonlocal thermodynamic formulation. The governing equations of the inhomogeneous continuum are obtained via the Hamilton principle. As a generalization of the constant order fractional continuum theory, the distributed order theory can model complex media characterized by inhomogeneous nonlocality and multiscale effects. In order to understand the correspondence between microscopic effects and the properties of the continuum, an equivalent mass-spring lattice model is also developed by direct discretization of the distributed order elastic continuum. Detailed theoretical arguments are provided to show the equivalence between the discrete and the continuum distributed order models in terms of internal nonlocal forces, potential energy distribution, and boundary conditions. These theoretical arguments facilitate the physical interpretation of the role played by the distributed order framework within nonlocal elasticity theories. They also highlight the outstanding potential and opportunities offered by this methodology to account for multiscale nonlocal effects. The capabilities of the methodology are also illustrated via a numerical study that highlights the excellent agreement between the displacement profiles and the total potential energy predicted by the two models under various order distributions. Remarkably, multiscale effects such as displacement distortion, material softening, and energy concentration are well captured at continuum level by the distributed order theory. 
\\

\noindent\keywords{Fractional calculus, Distributed order operators, Nonlocal elasticity theory, Nonlocal discrete lattices, Multiscale elasticity}
\end{abstract}
%%%%%%%%%%%%%%%%%%%%%%%%%%%%%%%%%%%%%%%%%%%%%%%%%%%%%%%%%%%%%%%%
\section{Introduction}
% Why should we study and model complex materials?
Recent developments seen in additive manufacturing technologies have rapidly changed the landscape of engineered materials by enabling the fabrication of complex architectured materials that were inconceivable only a decade ago. These novel materials find important applications in the most diverse fields of engineering, nanotechnology, biotechnology, and even medicine. Specific examples of the different applications of complex architectured materials include, but are not limited to, biological implants \cite{wang2016topological,yuan2019additive}, lightweight aerospace, automobile and naval structures \cite{marzocca2011review,romanoff2020review}, metamaterials designed for wave-guiding and vibration control \cite{hollkamp2020application,zhu2020nonlocal,nair2019nonlocal}, and even micro/nano-electromechanical devices \cite{ghorbani2013strain,iatsunskyi2015study}. In all these applications, the ability to achieve accurate predictions is paramount to deliver optimal designs and performance once these complex systems are deployed in the field. The need for accuracy combined with the rapidly increasing complexity of architectured materials continued strengthening the demand for modeling techniques capable of capturing the complex nature of these material systems.

%What is multiscale nonlocal elasticity and what are its key features?
In recent years, many experimental and theoretical investigations have reinforced the understanding that many of the above mentioned material classes exhibit non-negligible multiscale behavior. Additionally, these studies have also highlighted the prominent role of size-dependent effects (also referred to as nonlocal effects) that initially were believed to be important only for micro and nanoscale structures. Nonlocal effects can originate from different sources. The most well-known and studied included long-range interatomic and intermolecular interactions as well as material heterogeneity at micro- and nano-scales \cite{eringen1972linear,eringen1983differential,wang2011mechanisms}. However, nonlocal interactions can also be generated at the macroscale by means of intentionally nonlocal designs \cite{zhu2020nonlocal} and medium heterogeneity \cite{bavzant2000size,romanoff2020review,patnaik2021variable,silling2014origin}. Since the source of the nonlocal effects are typically specific to a material scale, the corresponding effects were often assumed to be localized at the respective scale while becoming negligible once integrated over the larger scales. This is one of the reason why early work on nonlocal scale effects was mostly focused on micro and nanoscale devices. However, in recent years the interest in nonlocal mechanics has rapidly expanded also following the realization that nonlocal forces naturally arises in the homogenization of heterogeneous systems, regardless of the scale. From these perspective, composite or porous materials can be seen as classical examples of nonlocal macroscale materials.

The understanding that nonlocality can arise at different scales, leads to an even more challenging scenario that is the possibility for nonlocal interactions to occur and interact across dissimilar scales. It is immediate to see how this latter scenario involves the simultaneous presence of nonlocal and multiscale elasticity concepts. This generalized elasticity problem involving nonlocal effects at different scales will be indicated in the rest of this manuscript as \textit{multiscale nonlocal elasticity}. The current literature of nonlocal elasticity has primarily been concerned with what might be referred to as single-scale nonlocality, and hence it has not specifically identified this multiscale nonlocal phenomenon. However, we emphasize that multiscale nonlocality should be expected in most real-world applications characterized by material and geometric heterogeneities. The latter conclusion follows not only from the observation of previously studied systems \cite{bavzant2000size,weinan2011principles,silling2010peridynamic}, but it also becomes a rather natural consequence of the design approach at the core of next generation architectured materials.

%What are the evidences or examples for multiscale nonlocal elasticity?
A few examples of multiscale nonlocal elastic behavior can be observed in real-world structures. Consider structures made from either functionally graded materials or porous materials with spatial gradations of the porosity. While the presence of medium heterogeneity (due to spatial variation in material properties or porosity) results in nonlocality \cite{patnaik2021variable}, the gradation of the structural properties, to (artificially) generate different underlying scales within the resulting structure, leads to a multiscale behavior \cite{weinan2011principles}. Multiscale nonlocal elasticity can also be found in other classes of structures such as, for example, structures made from layered composites \cite{liu2018multiscale}, elastic metamaterials \cite{zhu2020nonlocal}, multi-layer graphene sheets \cite{nazemnezhad2014free}, semiconductor devices containing wafers of different transition metal elements \cite{ghorbani2013strain}, and even electronic devices containing atomic coatings with different atoms \cite{iatsunskyi2015study}. In these different classes of structures, multiscale nonlocal elasticity can give rise to a variety of effects including, but not limited to, softening or stiffening, displacement distortion, anomalous wavenumber-frequency dispersion, energy concentration at different scales, and surface effects. The above discussion suggests that theoretical and numerical methodologies capable of capturing the simultaneous effect of nonlocality and multiscale behavior as well as their complex interaction will play an increasingly growing role to enable the design and performance prediction of the next generation of structures and materials.

%What approaches are currently available in literature to model multiscale nonlocal elasticity? What are their limitations?
\subsection{Brief overview of existing multiscale approaches}
Over the past few decades, several theoretical and computational approaches have been proposed to model the multiscale response of complex systems. Based on the underlying physical model, the most established multiscale approaches can be broadly classified into three categories: molecular approaches, local continuum approaches, and nonlocal continuum approaches. The following section is not meant to provide a comprehensive overview of such a broad and complex topic as multiscale mechanics, but it is intended to highlight some key aspects that limit the applications of established computational mechanics methodologies to the present problem of multiscale nonlocal elasticity. The key highlights of the different approaches and the challenges faced by them are summarized in the following:
\begin{itemize}
    \item \textit{Molecular approaches}: resolve the structure at the molecular level which, in turns, allows capturing fine level interactions either at the molecular or higher scales. Typical examples in this category include density functional theory~\cite{rasuli2010mechanical} and molecular dynamics~\cite{rapaport2004art}. By simultaneously accounting for the finest material scales and for all molecular interactions, molecular approaches can accurately capture multiscale nonlocal behavior. However, this class of techniques involves a number of degrees of freedom that scales proportionally to the number of particles, hence not making the approach suitable for simulations at the macro scales.
    
    \item \textit{Local continuum approaches}: leverage the classical (local) elastodynamic theory to simulate and predict the response of complex structures. Typical examples include discrete methods \cite{weinan2011principles} and asymptotic methods \cite{liu2018multiscale,craster2010high}. Discrete approaches such as, for example, finite element method, finite difference method, and model-order reduction techniques have found good success but are subject to an implicit trade-off between accuracy (directly related to the resolution of the specific discrete method) and computational time. In fact, for media with multiscale inhomogeneities (e.g. porous and fractal media) the discretization process requires fine spatial and temporal resolutions that lead rapidly to unattainable computational resources \cite{weinan2011principles,patnaik2021variable}. On the other hand, asymptotic methods, that use multiscale expansions of bulk (homogenized) material properties to capture information across scales, are often associated with lower computational costs. However, a major limitation of asymptotic methods follows from their rather complex analytical derivations that are only possible for limited types of structural analysis (and under specific loading conditions) \cite{craster2010high}. Finally, note that, irrespective of the nature of the method, the underlying classical continuum mechanics assumptions do not allow the resulting formulation to capture nonlocal effects at any given scale.
    
    \item \textit{Nonlocal continuum approaches}: extend the previous continuum formulations by introducing the contributions of long-range nonlocal interactions via differ-integral or integral constitutive equations. Depending on whether the nonlocal contributions are modeled using the strain field, the stress field, or the displacement field, the differ-integral approaches can be classified as strain-driven \cite{eringen1972nonlocal,polizzotto2001nonlocal}, stress-driven \cite{romano2017nonlocal}, or displacement-driven \cite{patnaik2022displacement,patnaik2020generalized}, respectively. Based on the nature of the kernels used to capture the nonlocal interactions, these approaches can be identified as integer-order \cite{eringen1972nonlocal,polizzotto2001nonlocal,romano2017nonlocal} (use exponential kernels) or fractional-order \cite{lazopoulos2006non,carpinteri2014nonlocal,sumelka2015fractional,patnaik2020generalized} (use power-law kernels typical of fractional calculus) approaches. Another rapidly growing approach to nonlocal mechanics is known as peridynamics \cite{silling2010peridynamic}; this is a purely integral method in nature. Discussions on the impact of differ-integral operators compared with purely integral operators are beyond the scope of this study. For more details, the interested reader is referred to \cite{silling2010peridynamic, patnaik2022displacement}. While existing classes of nonlocal approaches have been able to address a multitude of aspects typical of the response of size-dependent nonlocal structures, the nonlocal effects in these approaches are assumed to be restricted to a specific scale, typically the continuum scale. Consequently, based on the underlying formulation, they can capture only softening or stiffening response but not both simultaneously (a key feature of multiscale effects). Very recently, a few studies \cite{lim2015higher,patnaik2020towards} have expanded existing nonlocal approaches to model multiscale effects. However, these latter methods accounted for effects across only two scales and did not provide a general framework to be extended to multiple scales.
\end{itemize}
%A brief statement on what we want to do, to keep the flow.
From the above discussion, it emerges that existing approaches are not theoretically and computationally equipped to provide a comprehensive account of the physical phenomena involved in multiscale nonlocal elasticity. The present study attempts to address this technical gap by leveraging the nonlocal properties and intrinsic multiscale capabilities of distributed order (DO) operators.

%What is distributed order fractional calculus?
Distributed order operators are a natural multiscale generalization of the concept of constant order (CO) fractional operators. While a CO fractional operator is a differ-integral operator with a power-law kernel defined at a specific constant order, DO operators integrate the fundamental power-law kernel (typical of CO operators) over an extend range of orders~\cite{caputo1995mean}. Given that the fundamental kernel of a CO operator is retained, DO operators automatically inherit their nonlocal properties. It follows that DO operators can naturally capture nonlocal behavior defined at multiple scales. The inherent multiscale and nonlocal (in time and/or space) nature of DO operators has found applications in the modeling of several complex systems such as, for example, viscoelastic systems with multiple relaxation times \cite{atanackovic2002generalized}, anomalous transport processes marked by the presence of multiple temporal and spatial scales~\cite{metzler2004restaurant}, and materials with complex microstructure that evolve with externally applied thermal and/or mechanical loads \cite{lorenzo2002variable}. A detailed review of the applications of DO fractional calculus (DO-FC) to the analysis of real-world multiscale systems can be found in \cite{ding2021applications}. 

\subsection{Objectives and contributions of the study}
%What we want to do in this study?
In the present study, we leverage the unique properties of DO-FC to model multiscale nonlocal elasticity. Broadly speaking, the stress-strain constitutive relations are reformulated by means of DO operators in order to achieve a positive-definite and well-posed continuum formulation that captures nonlocal effects coexisting over an extended range of scales. The resulting theory can be interpreted as a fundamental extension to the existing continuum level approaches, because conceptually it combines the strengths of asymptotic (local continuum) and nonlocal continuum approaches. In fact, the DO model can be divided into two fundamental modules that capture the two different physical phenomena underlying multiscale nonlocal elasticity. First, we model nonlocal effects at a specific scale using a nonlocal continuum approach based on constant-order fractional mechanics. Next, we use a multiple scale expansion of the fractional-order via a distributed range of orders, analogous to asymptotic methods, to capture multiscale effects. 

The overall contributions of this study are four fold. 
\begin{itemize}
    \item We develop a generalized distributed order nonlocal elasticity theory (DO-NET) using DO-FC. Starting from a nonlocal thermodynamic formulation based on DO operators, DO fractional stress-strain constitutive relations are derived. Finally, the strong-form of the governing equations and of the boundary conditions is obtained using Hamilton's principle and standard variational simplifications. 
    
    \item We develop an equivalent mass spring lattice model (MSLM) by discretizing the DO-NET continuum governing equations and the associated boundary conditions. The MSLM is instrumental to obtain critical insights on the way distributed-order fractional operators can be leveraged to capture nonlocal interactions building up across scales. For this purpose, we transform the fundamental differ-integral operators to a purely integral form \cite{carpinteri2014nonlocal}. This allows the displacement derivatives within the DO operators of the stress-strain constitutive relation to be expressed in terms of the relative displacement of pairs of particles interacting at a particular scale. Upon discretization, the relative displacement terms can be interpreted as the elongation of elastic springs, enabling a straightforward (and transparent) route to derive the equivalent MSLM. 
    
    \item We complement the derivation of the MSLM from the strong-form governing equations by conducting two additional (theoretical) analyses that establish equivalence between the traction boundary condition and the potential energy of the discrete and continuum representations. The direct equivalence of the traction boundary conditions across the two representations (without requiring any constraints or additional terms) indicates a consistent model that is free from boundary and loading effects. In this regard, we note that the strain-driven and the stress-driven nonlocal continuum approaches require either additional (artificial) boundary constraints or correction terms (that depend on the specific boundary and loading conditions) in order to achieve well-posed continuum descriptions consistent with discretized lattice models \cite{challamel2014nonconservativeness}. Further, the analysis of equivalence of the potential energy across the two different representations helps identifying and isolating surface effects that are typical of multiscale nonlocal responses \cite{li2020contribution,wang2011vibration}.
    
    \item We conduct a comprehensive parametric study by simulating the response of a multiscale nonlocal structure. Simulations are performed via the MSLM and the DO-NET and evaluate the displacement field and the potential energy under different loading and boundary conditions, and order distributions. Apart from providing numerical evidence of the equivalence between the DO-NET and the MSLM (as expected), the parametric studies highlight the consistent predictions (free from boundary and loading effects) made from either approach. More importantly, the numerical studies provide critical insights on the impact of the nature of the order distribution on the (overall) degree of nonlocality of the structure. These qualitative insights are anticipated to aid identification of the order distribution in practical applications. Finally, we also use the numerical results to highlight that multiscale nonlocal effects such as displacement distortion, material softening, and energy concentration are well captured at the continuum level description of the distributed order theory.
\end{itemize}

% Structure of this paper
The remainder of this paper is structured as follows. In \S\ref{sec: 2}, we introduce a sample problem to illustrate the multiscale nonlocal behavior and the fundamental multiscale characteristic of DO operators. In \S\ref{sec: 3}, we develop the generalized DO-NET starting from nonlocal thermodynamic arguments. Next, in \S\ref{sec: 4}, we develop a physically consistent 1D MSLM starting from DO equilibrium relations derived in \S\ref{sec: 3}, and prove (theoretically) the force and energy equivalence between MSLM and DO-NET. Finally, in \S\ref{sec: 5} we numerically establish the equivalence between the MSLM and DO-NET via direct comparison of the mechanical responses obtained from both the approaches.

\section{The role of DO nonlocal elasticity theory}\label{sec: 2}
\begin{figure}[ht!]
	\centering
	\includegraphics[width=0.8\textwidth]{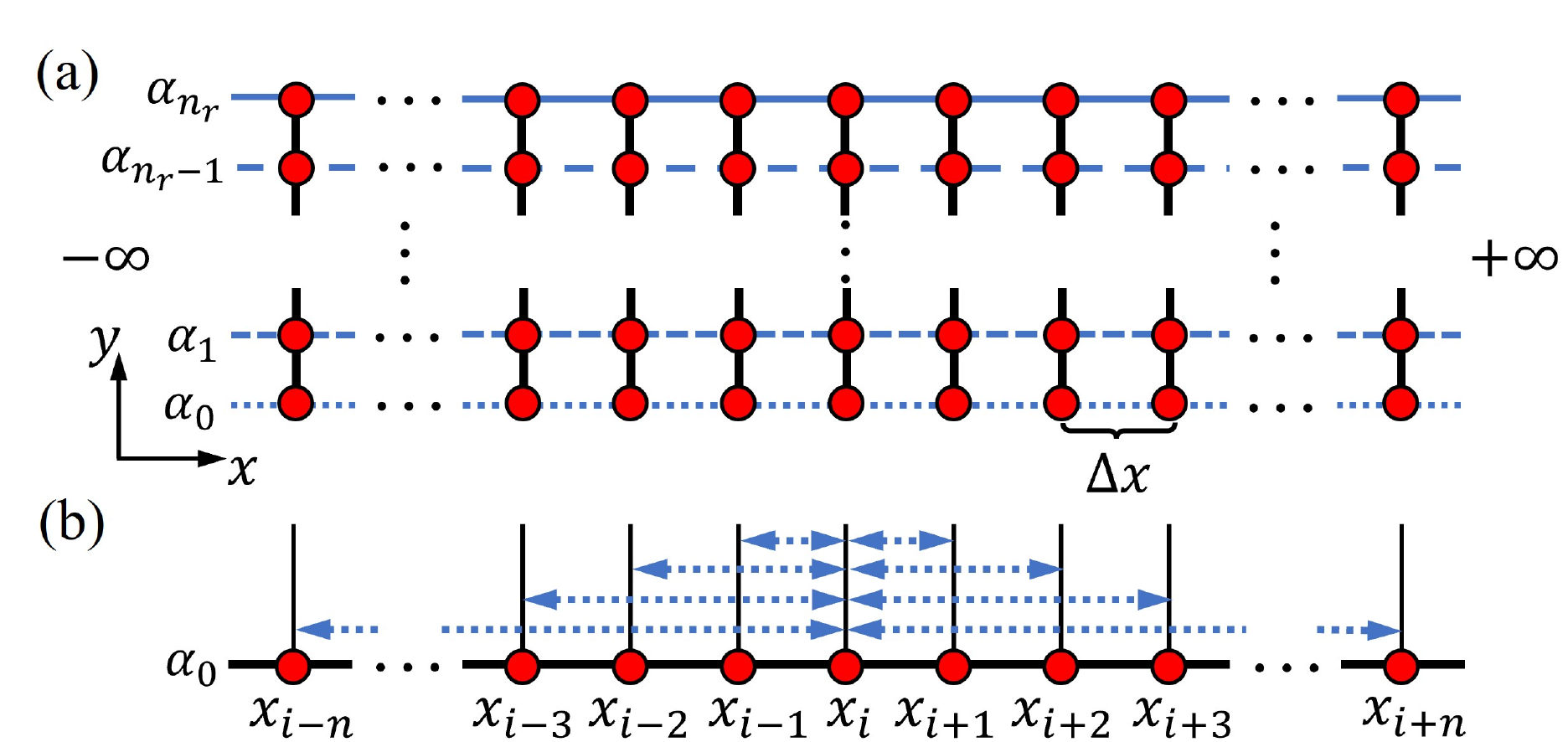}
	\caption{Schematic of the nonlocal lattice with transverse nonlocality. Each layer oriented along the $x$-direction can exhibit a different level of nonlocality or, equivalently, a different value of the exponent $\alpha$ of the power-law. (a) The mass points are periodically distributed in both the $x$ and $y$ directions. Dashed lines (aligned in the $x$ direction) represent the existence of nonlocal interactions between the masses on the layer. The nonlocal interaction is captured via a constant fractional order $\alpha_r$, $r=0,1,...,n_r-1,n_r$. Different dashed line styles represent different levels of nonlocality, hence different values of the constant order $\alpha_r$ along the $y$ direction. Solid lines (aligned in the $y$ direction) represent rigid connections between the point masses. (b) A closer view of the nonlocal interactions in a layer (e.g. $r=0$) and at a generic point $x_i$. The nonlocal interactions between the mass $x_i$ and the other lattice points are indicated by the dashed double-arrow lines.}
	\label{fig: DO_system}
\end{figure}

In this section, we will address the unique potential that DO operators offer with respect to modeling the response of multiscale nonlocal structures. To facilitate the understanding of how the DO-FC can enable multiscale nonlocal simulations, we discuss an illustrative example consisting of a two-dimensional transversely-nonlocal lattice. The 2D lattice can be seen as obtained by stacking a sequence of one-dimensional infinite nonlocal periodic lattices along the $y$-direction, as shown in Fig.~(\ref{fig: DO_system}). The 1D lattices are simplified microscopic representations of molecular chains (i.e. mass-spring chains typical of molecular dynamic formulations) and are subject to (nonlocal) long-range forces acting in the $x$-direction. Each layer can have a different degree of nonlocality, as schematically indicated by the different order $\alpha(y)$. It is this latter characteristic that determines its transversely nonlocal behavior.

The modeling of the above described lattice system builds upon two fundamental assumptions: 1) the strength of long-range interactions in each layer can be modeled via a power-law kernel of order $\alpha_r \in (0,1)$ with $r=0,1,...,m-1,m$, and 2) the different layers within the lattice are connected in the $y$-direction by stiff links (when compared to the more compliant nonlocal links in each layer). A detailed discussion on the motivation and physical significance of these assumptions is provided in the following:
\begin{itemize}
\item \textit{Assumption 1}: the assumption of power-law type long-range forces is motivated by early studies on nonlocal elasticity where different functional definitions of the nonlocal kernel were obtained by matching interatomic nonlocal behaviors~\cite{eringen1972linear,eringen1983differential}. Following a similar conceptual approach, the order $\alpha_r$ (which characterizes the strength of the nonlocal interactions) in each layer can be obtained by fitting the power-law kernel against experimentally or theoretically derived long-range interaction data. The selection of the power-law kernels allows capturing the nonlocal behavior of a given layer by means of constant-order fractional continuum models \cite{patnaik2020towards,carpinteri2014nonlocal}. It follows that this assumption does not limit the general character of the modeling approach and of the results, while being justified by available experimental evidence. We merely note that recent studies~\cite{patnaik2022displacement} have also presented nonlocal formulations capable of using generalized attenuation kernels. These results suggest that future extension of the DO approach to general kernels could also be envisioned. Following the power-law assumption, the force of interaction between two particles $p$ and $q$ (where $p\neq q$) located on the layer $r$ can be expressed as:
\begin{equation}
\label{eq: DO_mot_int_f}
	F_{pq}^{(r)}(x_p,x_q,r)=\frac{F_{0}^{r} \Delta u_{pq}}{|x_p-x_q|^{2+\alpha_r}} 
\end{equation}
where $F_{0}^{(r)}$ can be interpreted as a material constant that carries information about the the elastic constants (analogous to local discrete MSLM models where the constant is typically referred to as the spring constant), the characteristic length~\cite{eringen1983differential}, and the strength of the nonlocal interactions within the layer. Further, $\Delta u_{pq} = u_p - u_q$ denotes the relative displacement between the two particles. Thus, in general, $F_0^r \triangleq F_0^r(\alpha_r)$.

\item \textit{Assumption 2}: the rigid connections in the $y$-direction guarantee that the motion of individual particles in each layer is restricted to the axial direction. We will show that this assumption allows for a simpler yet comprehensive derivation of a DO operator that evidently characterizes the response of the overall lattice. Note that this assumption of negligible through-the-thickness deformation is typical of many structural applications involving slender structures (e.g. beams and plates).
\end{itemize}

We consider the problem of homogenizing the 2D lattice to achieve an equivalent 1D nonlocal continuum capable of accurately capturing the axial motion of the medium subject to a transverse distribution of nonlocality (see Fig.~(\ref{fig: MSLM})). For this purpose, we can separate the process into two steps: 1) reducing (i.e. homogenizing) the initial 2D lattice to a 1D lattice oriented along the $x$-direction (hence collapsing the $y$-dimension), and 2) taking the continuum limit of the resulting 1D lattice to obtain the final 1D continuum.

In the first step, the transverse dimension $y$ is reduced (e.g. assuming that the characteristic dimension in the $y$-direction is negligible compared to that in the $x$-direction) so that the nonlocal interactions between two particles $p$ and $q$ on any given layer can be expressed as a summation of the contributions from each layer (conceptually equivalent to systems of springs in parallel), in the following manner:
\begin{equation}
\label{eq: DO_hom_exp}
	F_{pq}^{(H)}(x_p,x_q)= \sum_{r=0}^{n_r}{F_{pq}^{(r)}}(x_p,x_q,r)
\end{equation} 
where the superscript $(H)$ denotes the nonlocal force in the homogenized 1D lattice. The total nonlocal force on the generic particle $p$ (due to all the remaining particles in the homogenized 1D infinite medium) can be obtained by using Eqs.~(\ref{eq: DO_mot_int_f},\ref{eq: DO_hom_exp}):
\begin{equation}
\label{eq: DO_hom_tot_F}
	F_{p}^{(H)} = \sum_{q=-\infty}^{q=\infty}F_{pq}^{(H)}(x_p,x_q) =\sum_{r=0}^{n_r}\sum_{q=-\infty}^{q=\infty}\frac{F_{0}^{r}(\alpha_r) \Delta u_{pq}}{|x_p-x_q|^{2+\alpha_r}}
\end{equation} 
The double summation in Eq.~(\ref{eq: DO_hom_tot_F}) (one over the number of layers in the transverse direction, and one over the number of particles in the axial direction) allows capturing the effect of the interatomic forces acting in parallel and it will be shown to be at the basis of the occurrence of DO operators in the homogenized continuum. To further substantiate this argument, consider the continuum limit expression for the total force $F_{p}^{(H)}$ in Eq.~(\ref{eq: DO_hom_tot_F}). For this purpose, we cast the inter-particle (discrete) force constant $F_{0}^{r}(\alpha_r)$ in the following manner:
\begin{equation}
    \label{eq: F_cont_expression}
    F_{0}^{r}(\alpha_r) = \underbrace{k_0 \Delta x}_{T_1} \underbrace{\kappa(\alpha_r) \Delta \alpha}_{T_2}
\end{equation}
where $k_0$ denotes a force per unit displacement (analogous to the stiffness constant of a discrete spring) of the nonlocal lattice, and $\Delta x$ denotes the periodic spacing along the axial direction of the lattice (see Fig.~(\ref{fig: DO_system})). In the context of the 2D lattice, these terms (combined as $T_1$) capture the discrete equivalent of the elastic properties of the 1D nonlocal continuum. Analogous to the power-law index\footnote{Note that the power-law index of functionally graded materials is different from the power-law index (or, equivalently, exponent) of the power-law kernel used in fractional-order approaches to nonlocal elasticity.} that controls material properties in functionally graded materials~\cite{marzocca2011review}, the remaining two terms (combined as $T_2$) capture the variation in the elastic properties (i.e. in $T_1$) along the $y$-direction in the original 2D lattice. The dimensionless function $\kappa(\alpha)$ physically denotes the change in material properties per unit order. This latter aspect is more evident considering the transformation $\alpha = g(y)$, where the function $g(y)$ captures the variation of the degree of nonlocality across different layers stacked in the transverse direction. Under this transformation, $T_2$ can be expressed as:
\begin{equation}
    T_2 = \kappa(\alpha_r)\Delta\alpha =\kappa(g(y_r)) \left[Dg(y)|_{y_r}\right] \Delta y
\end{equation}
where $y_r$ is the spatial location along the transverse direction such that $\alpha_r = g(y_r)$ and $D(\cdot)$ denotes the first order spatial derivative. Combining the functions within the composite function $\kappa(g(y_r))$, followed by a product and grouping of $\kappa(g(y_r))$ and $\left[Dg(y)|_{y_r}\right]$, the above expression can be recast in the following manner:
\begin{equation}
    T_2 = \kappa(\alpha_r)\Delta\alpha =\widetilde{\kappa}(y_r) \Delta y
\end{equation}
It follows that the function $\widetilde{\kappa}(y)$ denotes the strength of the variation of the degree of nonlocality across the transverse direction, which is expressed analogously via $\kappa(\alpha)$, albeit through the order variable.

Now, by substituting the expression for $F_0^r$ in Eq.~(\ref{eq: DO_hom_tot_F}) and taking the continuum limit, we obtain:
\begin{equation}
\label{eq: DO_hom_tot_F_cont_step3}
	\lim_{\Delta \alpha \rightarrow 0} \left[ \lim_{\Delta x \rightarrow 0} F_{p}^{(H)} \right] =  k_0 \underbrace{\int_{\alpha_\text{min}}^{\alpha_\text{max}} \kappa(\alpha) \underbrace{\left[ \int_{-\infty}^{\infty} \frac{u(x_p)-u(x^\prime)}{|x_p-x^\prime|^{2+\alpha}} \mathrm{d} x^\prime \right]}_\text{CO fractional derivative} \mathrm{d}\alpha}_\text{DO fractional derivative}
\end{equation}
where $\alpha_{min}$ and $\alpha_{max}$ are the minimum and maximum value of constant-order $\alpha_r$. As evident from the above expression, DO derivatives appear naturally from the homogenization process applied to a complex medium characterized by heterogeneous distribution of nonlocal effects. In other terms, the transverse stacking of the nonlocal chains with different degree of nonlocality generates different material scales within the structure which are captured within the order variation of the DO derivative. It is important to note that, the order variation can be localized to a single material scale such as, for example, porous beams with spatially varying degree of porosity (see \cite{patnaik2021variable}). However, in this latter case, the material is heterogeneous in nature with properties (e.g. in the case of a porous beams, properties could refer to the level of porosity) localized to a specific material scale. We merely note that this class of structures are better described by a different class of fractional operators denominated variable-order operators \cite{patnaik2021variable}. In the present case (Fig.~(\ref{fig: DO_system})), the material properties (or particles) are uniform and localized on a specific layer. Additionally, each layer exhibits nonlocal interactions of a specific degree that is localized on the given layer. Hence, the existence of the order variation is a manifestation of the multiscale nature of the structure and not merely of the material inhomogeneity localized to a specific scale. In fact, a further look at Eq.~(\ref{eq: F_cont_expression}) and Eq.~(\ref{eq: DO_hom_tot_F_cont_step3}) highlights how the multiple scales and the material properties localized at the corresponding scales are captured via the strength function of the DO derivative. In summary, the application of DO operators is suitable for structures where the order variation is resolved across different (coexisting) material scales, while the application of variable-order operators is suitable when the order variation is localized (or resolved) within a single material scale. In very crude terms, this latter concept can be interpreted as analogous to material scales coexisting in a parallel (DO) or series (VO).

The above discussion provides a simple yet powerful physical interpretation of the DO operator for mechanics and create a route to understand and model continua of practical interest such as, for example, multi-layer graphene sheets~\cite{nazemnezhad2014free}, multi-layer heterogeneous atomic coatings~\cite{iatsunskyi2015study}, semiconductor devices containing wafers of different elements~\cite{ghorbani2013strain}, and other multiscale structures mentioned previously in the introduction. This section illustrated, via a simple example, the multiscale origin of DO operators in elastic continua. In the following, we will develop a complete and rigorous formulation of three-dimensional DO elasticity.

\section{Mathematical formulation of DO nonlocal elasticity}\label{sec: 3}
In this section, we formulate the DO nonlocal elasticity theory (DO-NET). Specifically, we develop the DO-NET framework by using a classical definition of the local kinematics and a generalized DO nonlocal thermodynamic framework. By satisfying the first and second principle of thermodynamics, an Eringen-like nonlocal definition of the DO stress can be obtained. Based on the kinematics and constitutive relations, the governing equations and the associated boundary conditions will be derived using both the Hamilton principle and the balance of linear momentum. Note that while there exist other alternative approaches to deriving DO theory (for example, developing the DO version of the existing CO formulations based on fractional kinematics~\cite{patnaik2020generalized,sumelka2014thermoelasticity}), in this study we take an approach based on nonlocal constitutive relations.

\subsection{Derivation of DO constitutive relations}\label{ssec: 3.1}
In analogy with seminal approaches to nonlocal elasticity based on nonlocal stress-strain constitutive relations~\cite{eringen1972nonlocal,eringen1983differential}, we start from the classical (local) definition of the deformation gradient:
\begin{equation}
    \bm{F}=\frac{\mathrm{d}\bm{x}}{\mathrm{d}\bm{X}}
\end{equation}
where $\bm{x}$ and $\bm{X}$ denote coordinates in the deformed and undeformed configuration, respectively. Recall that the deformation gradient tensor $\bm{F}$ relates differential line elements $\mathrm{d}\bm{x}$ and $\mathrm{d}\bm{X}$ within the deformed and undeformed configurations, so that the infinitesimal strain tensor can be expressed as: 
\begin{equation}
    \bm{\epsilon}=\frac{1}{2}\left(\bm{\nabla}\bm{u}+\bm{\nabla}^{\bm{T}}\bm{u}\right)
\end{equation}
where $\bm{u}=\bm{x}-\bm{X}$ represents the displacement field, and $\bm{\nabla}(\cdot)$ is the gradient operator. 

By using the above described kinematic relations in conjunction with thermodynamic equilibrium conditions, nonlocal constitutive relations can be obtained. For a nonlocal solid, the energy at a given point is affected by long range cohesive interactions (hence by energy exchange) with other particles within the horizon of nonlocality. Consequently, the internal energy density is a functional such that $e=e(\bm{\epsilon},\mathscr{R}(\bm{\epsilon}),\eta)$ \cite{polizzotto2001nonlocal}. In this functional, $\mathscr{R}(\cdot)$ is a linear integral operator that captures the nonlocal energy exchanges, and $\eta$ is the entropy. In classical nonlocal approaches, these linear integral operators $\mathscr{R}(\cdot)$ are defined using monotonically decaying kernels such as, for example, exponential kernels \cite{polizzotto2001nonlocal} or CO power-law kernels \cite{carpinteri2014nonlocal}. As an example, Carpinteri's 1D CO nonlocal elasticity formulation~\cite{carpinteri2014nonlocal}, the linear integral operator that constitutes the energy density as well as the constitutive relation is defined using a CO fractional integral as:
\begin{equation}\label{eq: 1D Carpinteri sigma}
    \sigma^{\mathrm{CO}}(x)=\mathscr{R}^{\mathrm{CO}}(\epsilon)=E\prescript{R-RL}{a}{}\bm{I}_b^{\alpha}\epsilon(x)
\end{equation}
where $E$ is the Young's modulus of the 1D solid, and the superscript $\square^{\mathrm{CO}}$ indicates constant order operators. Although the R-RL linear integral operator in Eq.~(\ref{eq: 1D Carpinteri sigma}) can capture nonlocality, its CO power-law attenuation kernel is not general enough to capture complex multiscale nonlocal effects. In the following, we will generalize this constitutive formulation to the DO form.

Consider a 3D nonlocal solid that exhibits multiscale nonlocal interactions similar to the 2D lattice structure shown in Fig.~(\ref{fig: DO_system}). We define the following 3D DO linear integral operator:
\begin{equation}\label{eq: DO linear integral operator}
\begin{aligned}
    \mathscr{R}(\bm{\epsilon})&=\prescript{R-RL}{a_1}{}\mathcal{I}_{b_1}^{1-\alpha,\bm{\kappa}(\alpha)}\prescript{R-RL}{a_2}{}\mathcal{I}_{b_2}^{1-\alpha,\bm{\kappa}(\alpha)}\prescript{R-RL}{a_3}{}\mathcal{I}_{b_3}^{1-\alpha,\bm{\kappa}(\alpha)}(\bm{C}:\bm{\epsilon})\\
    &=\int_{0}^{1}\kappa_{ijkl}(\alpha)\left(\prescript{R-RL}{a_1}{}\bm{I}_{b_1}^{1-\alpha}\prescript{R-RL}{a_2}{}\bm{I}_{b_2}^{1-\alpha}\prescript{R-RL}{a_3}{}\bm{I}_{b_3}^{1-\alpha}\right)\left(C_{ijkl}\epsilon_{kl}\right)\textrm{d}\alpha\\
\end{aligned}
\end{equation}
where $\bm{\kappa}(\alpha)=\kappa_{ijkl}(\alpha)$ is the fourth-order tensorial strength-function. The product of the two fourth-order tensors ($\bm{\kappa}$ and $\bm{C}$) via the subscripts $\{i,j,k,l\}$ follows the Hadamard product rule. Subscripts $\{1,2,3\}$ represent the three orthonormal axes in 3D space. Further, $\prescript{R-RL}{a_i}{}\mathcal{I}_{b_i}^{1-\alpha,{\kappa}(\alpha)}$ (where $i\in\{1,2,3\}$) are the DO Riesz Riemann-Liouville integrals introduced in Eq.~(S7). In contrast to the CO operator in Eq.~(\ref{eq: 1D Carpinteri sigma}), which only captures nonlocality characterized by a CO strength $\alpha$, the proposed DO linear integral operator serves as a superposition of different orders that allows capturing multiscale nonlocality (see discussion on DO operators in SM~\S{S1} and \S\ref{sec: 2}).

According to Polizzotto~\cite{polizzotto2001nonlocal}, the first and second principle of thermodynamics are modified for a nonlocal elastic solid as:
\begin{subequations}\label{eq: nonlocal thermodynamics-1}
\begin{align}
    \dot{e}&=\bm{\sigma}:\dot{\bm{\epsilon}}+h-\nabla{\cdot}\bm{q}+P \quad &&\forall \bm{x} \in \bm{V}\\
    T\dot{\eta}_{int}&=\bm{\sigma}:\dot{\bm{\epsilon}}-\dot{\psi}-\eta\dot{T}-{\nabla}T\cdot\frac{\bm{q}}{T}+P \geq 0 \quad &&\forall \bm{x} \in \bm{V}
\end{align}
\end{subequations}
where $\dot{\square}$ denotes the first-order time derivative, $h$ is the heat generated internally per unit volume, $\bm{q}$ is the heat flux density, $T$ is the absolute temperature, $\psi$ is the Helmholtz free energy, and $\bm{V}$ is the total volume of the solid. The above thermodynamic balance laws differ from a classical form by the term $P$ on their right-hand side. The term $P$ in the above expressions is added to enable a point-wise (strong) enforcement of the thermodynamic balance laws for nonlocal solids \cite{polizzotto2001nonlocal}. As discussed in \cite{polizzotto2001nonlocal}, $P$ can be interpreted as a nonlocal energy residual that represents the energy exchanged by a point with all other points within its nonlocal horizon. Although Eqs.~(\ref{eq: nonlocal thermodynamics-1}) differ from the classical first and second principle of thermodynamics due to the term $P$, the weak form of these equations is obtained as:
\begin{subequations}\label{eq: nonlocal thermodynamics-2}
\begin{align}
    \int_{\bm{V}}\dot{e}\mathrm{d}\bm{V}&=\int_{\bm{V}}\left(\bm{\sigma}:\dot{\bm{\epsilon}}+h-\nabla{\cdot}\bm{q}\right)\mathrm{d}\bm{V} \quad &&\forall \bm{x} \in \bm{V}\\
    \int_{\bm{V}}\dot{\eta}_{int}\mathrm{d}\bm{V}&=\int_{\bm{V}}\left(\bm{\sigma}:\dot{\bm{\epsilon}}-\dot{\psi}-\eta\dot{T}-{\nabla}T\cdot\frac{\bm{q}}{T}\right)\mathrm{d}\bm{V} \geq 0 \quad &&\forall \bm{x} \in \bm{V}
\end{align}
\end{subequations}
which are identical to their classical counterparts. Note that the nonlocal energy residual $P$ vanishes under integration because of the insulation condition (see \cite{polizzotto2001nonlocal}).

Using the weak statement of thermodynamics shown in Eq.~(\ref{eq: nonlocal thermodynamics-2}-b), we derive the nonlocal constitutive relations. Since Eq.~(\ref{eq: nonlocal thermodynamics-2}-b) must hold for any thermo-elastic deformation process, we first consider thermo-elastic deformation at uniform temperature, that is ${\nabla}T=0$ within the volume $\bm{V}$. Next, recall that for local solids the Helmholtz free energy has functional dependence on the local strain and temperature, that is, $\psi=\psi(\bm{\epsilon,T})$. However, for nonlocal solids, the additional nonlocal interactions alter this functional dependence and we have $\psi=\psi(\bm{\epsilon},\mathscr{R}(\bm{\epsilon)},T)$, where the additional functional dependence of $\psi$ on the integral operator $\mathscr{R}$ accounts for the impact of the nonlocal interactions. Finally, by substituting $\nabla T=0$ and the functional relationship of the Helmholtz free energy $\psi=\psi(\bm{\epsilon},\mathscr{R}(\bm{\epsilon)},T)$ into Eq.~(\ref{eq: nonlocal thermodynamics-2}-b) we obtain \cite{polizzotto2001nonlocal}:
\begin{equation}\label{eq: second nonlocal thermodynamics}
    \int_{\bm{V}}\left(\sigma - \frac{\partial\psi}{\partial\bm{\epsilon}} - \mathscr{R}\left(\frac{\partial\psi}{\partial\mathscr{R}(\bm{\epsilon})}\right)\right):\dot{\bm{\epsilon}}\mathrm{d}\bm{V}
    - \int_{\bm{V}}\left(\eta + \frac{\partial\psi}{\partial{T}}\right)\dot{T}\mathrm{d}\bm{V} \geq 0 \quad \forall \bm{x} \in \bm{V}
\end{equation}
To guarantee the inequality holds for any thermo-elastic process with arbitrary $\dot{\bm{\epsilon}}$ and $\dot{T}$, we obtain the following constitutive laws:
\begin{subequations}\label{eq: constitutive laws}
\begin{align}
    \bm{\sigma}&=\frac{\partial\psi}{\partial\bm{\epsilon}} + \mathscr{R}\left(\frac{\partial\psi}{\partial\mathscr{R}(\bm{\epsilon})}\right) &&\forall \bm{x} \in \bm{V}\\
    \eta&=-\frac{\partial\psi}{\partial{T}} &&\forall \bm{x} \in \bm{V}
\end{align}
\end{subequations}
Substituting them back into Eq.~(\ref{eq: nonlocal thermodynamics-1}-b), we obtain the strong statement of the second law of thermodynamics:
\begin{equation}
    T\dot{\eta}_{int}=-{\nabla}T\cdot\frac{\bm{q}}{T} \geq 0, \quad \forall \bm{x} \in \bm{V}\\
\end{equation}
Eq.~(\ref{eq: constitutive laws}-a) provides a thermodynamic restriction to the constitutive equations for nonlocal elasticity. Finally, we define the Helmholtz free energy as:
\begin{equation}\label{eq: Helmholtz free energy}
    \psi=\frac{1}{2}\bm{\epsilon}:\mathscr{R}(\bm{\epsilon})
\end{equation}
Substituting the above expression within Eq.~(\ref{eq: constitutive laws}-a), we obtain the DO nonlocal constitutive relation (and its indicial form) as:
\begin{equation}\label{eq: DO constitutive relation}
\begin{aligned}
    \bm{\sigma}&=\mathscr{R}(\bm{\epsilon})=\prescript{R-RL}{a_1}{}\mathcal{I}_{b_1}^{1-\alpha,\bm{\kappa}(\alpha)}\prescript{R-RL}{a_2}{}\mathcal{I}_{b_2}^{1-\alpha,\bm{\kappa}(\alpha)}\prescript{R-RL}{a_3}{}\mathcal{I}_{b_3}^{1-\alpha,\bm{\kappa}(\alpha)}(\bm{C}:\bm{\epsilon})\\
    \sigma_{ij}&=\int_{0}^{1}\kappa_{ijkl}(\alpha)\left(\prescript{R-RL}{a_1}{}\bm{I}_{b_1}^{1-\alpha}\prescript{R-RL}{a_2}{}\bm{I}_{b_2}^{1-\alpha}\prescript{R-RL}{a_3}{}\bm{I}_{b_3}^{1-\alpha}\right)\left(C_{ijkl}\epsilon_{kl}\right)\textrm{d}\alpha\\
\end{aligned}
\end{equation}
Note that, similar to the stiffness tensor $\bm{C}$, the symmetry of the strain and stress tensors requires that the strength-function tensor $\bm{\kappa}$ satisfies both major and minor symmetries. A detailed proof of this claim is provided in SM \S{S4}.

By using the definitions of the stress and strain fields provided above, the total deformation energy $(\Pi)$ of the nonlocal solid can be expressed as:
\begin{equation}\label{eq: Explicit potential energy}
    \Pi=\frac{1}{2}\int_{\bm{V}}\bm{\sigma}:\bm{\epsilon}\mathrm{d}\bm{V}
    =\frac{1}{2}\int_{\bm{V}}\bm{\epsilon}:\prescript{R-RL}{a_1}{}\mathcal{I}_{b_1}^{1-\alpha,{\bm{\kappa}}(\alpha)}\prescript{R-RL}{a_2}{}\mathcal{I}_{b_2}^{1-\alpha,{\bm{\kappa}}(\alpha)}\prescript{R-RL}{a_3}{}\mathcal{I}_{b_3}^{1-\alpha,{\bm{\kappa}}(\alpha)}(\bm{C}:\bm{\epsilon})\mathrm{d}\bm{V}
\end{equation}
Note that, unlike the local continuum description, the potential energy obtained following the DO approach is not quadratic in nature. In order to ensure that the potential energy is positive-definite and that the governing equations derived from the defined potential energy are well-posed, it is required that the attenuation kernel is positive-definite and symmetric in nature \cite{polizzotto2001nonlocal}. In this regard, the DO formulation presented in Eq.~(\ref{eq: DO constitutive relation}) does not violate these requirements since the DO operator admits the positive-definite and symmetric power-law kernel within the individual CO fractional integrals ($\prescript{R-RL}{a_i}{}\bm{I}_{b_i}^{1-\alpha}(\cdot)$). 

Although the condition of symmetry helps achieving a positive-definite and well-posed formulations, in classical (integer-order) approaches to nonlocal elasticity this condition restricts the application of the resulting theory to isotropic structures; hence, leading to a modeling approach not general enough to represent many applications of practical interest and characterized by asymmetric interactions \cite{silling2010peridynamic, hollkamp2020application, patnaik2022displacement}. This aspect was also very recently discussed in \cite{batra159misuse} which highlighted the misuse of a classical Eringen's nonlocal integral approach in functionally graded materials.
In this context, note that, although the DO formulation proposed in Eq.~(\ref{eq: DO constitutive relation}) adopts a symmetric power-law kernel, unlike CO operators, the tensorial strength-function $\bm{\kappa}(\alpha)$ (see Eq.~(\ref{eq: DO linear integral operator})) allows the DO formulation to easily capture anisotropic nonlocality as well as material heterogeneity. On the other hand, the strength-function serves as an additional parameter (beyond the size of the nonlocal horizon, and the fractional-order) that allows combining the different fractional integrals in a 'heterogeneous' fashion by weighting them using order-dependent strengths (that is, $\kappa_{ijkl}(\alpha)$). In other terms, the strength-function can be tuned to model material heterogeneity while still adopting a symmetric kernel and hence, a positive-definite formulation.

In order to better illustrate the capability of the DO formulation to model anisotropic nonlocality as well as material heterogeneity, consider the following DO nonlocal formulation applied to anisotropic nonlocal materials:
\begin{equation}\label{eq: sigma_anisotropic_heterogeneous}
    \sigma_{ij}(\bm{x})=\mathbb{I}^{1-\alpha,\bm{\kappa}(\alpha)}\left(\lambda(\bm{x})\delta_{ij}\epsilon_{kk}+2\mu(\bm{x})\epsilon_{ij}\right)
\end{equation}
with spatial-dependent heterogeneous Lam\'e parameters $\mu(\bm{x})$ and $\lambda(\bm{x})$. Note that here we use the compact notation $\mathbb{I}^{(\cdot)}$ to denote the 3D DO nonlocal integral in Eq.~(\ref{eq: DO linear integral operator}). To show how DO-NET can explicitly model anisotropic nonlocality and material heterogeneity, we define the component of the strength-function tensor as:
\begin{equation}\label{eq: kappa_anisotropic}
\begin{aligned}
    \kappa_{11}=\kappa_{22}=\kappa_{33}=\kappa_1(\alpha),\quad \kappa_{12}=\kappa_{13}=\kappa_{23}=\kappa_2(\alpha),\quad \kappa_{44}=\kappa_{55}=\kappa_{66}=\kappa_3(\alpha)\\
\end{aligned}
\end{equation}
such that the corresponding stress components obtained via DO-NET in Eq.~(\ref{eq: sigma_anisotropic_heterogeneous}) can be further expressed as:
\begin{equation}\label{eq: sigma_anisotropic_heterogeneous_1}
\begin{aligned}
    \sigma_{ij}&=(2\overline{\mu}+\overline{\lambda})\underbrace{\mathbb{I}^{1-\alpha,\kappa_{1}(\alpha)}\epsilon_{ij}}_{\textrm{axial nonlocal effects}} + \overline{\lambda}\underbrace{\mathbb{I}^{1-\alpha,\kappa_{2}(\alpha)}\delta_{ij}(\epsilon_{kk}-\epsilon_{ij})}_{\textrm{lateral nonlocal effects}},&& i=j\\
    \sigma_{ij}&=2\overline{\mu}\underbrace{\mathbb{I}^{1-\alpha,\kappa_{3}(\alpha)}\epsilon_{ij}}_{\textrm{shear nonlocal effects}},&& i~{\neq}~j\\
\end{aligned}
\end{equation}
It is seen that, by reformulating the DO stress in Eq.~(\ref{eq: sigma_anisotropic_heterogeneous}), we explicitly obtain the contributions of the nonlocal effects brought by the anisotropic deformation and accordingly define the effective Lam\'e parameters $\overline{\mu}$ and $\overline{\lambda}$ to account for material heterogeneity.
By observing the DO constitutive relation in Eq.~(\ref{eq: sigma_anisotropic_heterogeneous_1}), we immediately obtain that the DO formulation can model anisotropy via the strength-tensor. Specifically, while traditional nonlocal theories (such as Eringen's nonlocal elasticity~\cite{eringen1972nonlocal}) use a single attenuation function to capture homogeneous nonlocal effects, $\kappa_1(\alpha)$, $\kappa_2(\alpha)$, and $\kappa_3(\alpha)$ can be tuned to capture anisotropic nonlocal effects caused by axial, lateral, and shear deformation, respectively. Moreover, unlike Eringen's formulation which can only model homogeneous materials to guarantee positive-definiteness and symmetry of attenuation kernel~\cite{polizzotto2001nonlocal}, the strength-function tensor ($\kappa_{ijkl}(\alpha)$) within the DO operator can be used to model material heterogeneity, while the inner attenuation kernel is still symmetric and hence satisfies the necessary (kernel) requirements. From a mathematical perspective, it appears that the DO formulation in Eq.~(\ref{eq: DO constitutive relation}) captures the heterogeneity in the nonlocal solid via the order-dependent strength-function instead of using the spatially-dependent (anisotropic) constitutive matrices. This is remarkable since the DO theory provides a possible route to model anisotropic nonlocality via nonlocal constitutive relations while still achieving a positive definite potential energy (through the symmetric power-law kernel). We merely note that the expressions in Eq.~(\ref{eq: sigma_anisotropic_heterogeneous_1}) suggest that the proposed formulation also compares closely with the anisotropic nonlocal theory proposed in~\cite{lazar2020three} where a so-called two-fold anisotropy is introduced using anisotropic nonlocal kernel functions.

\subsection{Derivation of Governing Equations via Hamilton's principle}\label{ssec: 3.2}
In this section, we derive the strong-form of the DO equilibrium equations governing the response of the nonlocal solid by using variational principles. Note that the governing equations could also be derived by using linear momentum balance over a representative volume element of the nonlocal domain (see SM~\S{S4}). Both approaches yield identical results, however the variational approach is more immediate while the use of linear momentum balance requires extra care in accounting for the effects of the nonlocal interactions \cite{polizzotto2001nonlocal, sidhardh2020thermodynamics}. 

We use the Hamilton's principle to derive the strong-form of the governing equations:
\begin{equation}
\label{eq: Hamilton}
    \delta\int_{t_0}^{t_1}(\mathcal{W}+\mathcal{T}-\Pi)\mathrm{d}t=0
\end{equation}
In the above equation, $\mathcal{T}$ and $\mathcal{W}$ denote the kinetic energy and the work done by externally applied forces on the nonlocal solid, respectively. Note that the nonlocality does not alter either the expressions of the kinetic energy or of the external work done on the solid. $\mathcal{T}$ and $\mathcal{W}$ are expressed as:
\begin{equation}
\begin{aligned}
    \mathcal{T}&=\frac{1}{2}\int_{\bm{V}}\rho\left(\bm{\dot{u}}\cdot\bm{\dot{u}}\right)\mathrm{d}\bm{V}\\
    \mathcal{W}&=\int_{\bm{V}}\rho\bm{b}\cdot\bm{u}\mathrm{d}\bm{V}+\int_{\bm{\Gamma}}\bm{t}{\cdot}\bm{u}\mathrm{d}{\bm{\Gamma}}\\
\end{aligned}
\end{equation}
where $\rho$ denotes the density of the solid, $\bm{b}$ denotes the volumetric body forces applied, and $\bm{t}$ denotes the surface tractions.
By substituting the expressions for the deformation energy, kinetic energy, and external work in the Hamilton's principle in Eq.~(\ref{eq: Hamilton}) and applying standard rules of variational calculus, we obtain:
\begin{equation}
\label{ref: EOM_simplification}
\begin{aligned}
    \int_{t_0}^{t_1}\left[\int_{\bm{V}}\left(\bm{\nabla}\cdot\bm{\sigma}+\rho\bm{b}-\rho\Ddot{\bm{u}}\right){\delta}\bm{u}\mathrm{d}\bm{V}+\int_{\bm{\Gamma}}\left(\bm{\sigma}\cdot\bm{n}-\bm{t}\right){\delta}\bm{u}\textrm{d}{\bm{\Gamma}}\right]\textrm{d}t=0\\
\end{aligned}
\end{equation}
Since the above expression must hold true for all variations $\delta\bm{u}$ and all possible time intervals $[t_0,t_1]$, we obtain the following governing equation: 
\begin{equation}\label{eq: EOM}
\begin{aligned}
    \bm{\nabla}\cdot\bm{\sigma}+\rho\bm{b}-\rho\bm{\Ddot{u}}&=0,\quad \forall \bm{x} \in \bm{V}\\
\end{aligned}
\end{equation}
subject to the following boundary conditions:
\begin{equation}
\begin{aligned}
    \bm{\sigma}\cdot\bm{n}-\bm{t}&=0, \quad \forall \bm{x} \in \bm{\Gamma}\\
    \delta\bm{u}&=0, \quad \forall \bm{x} \in \bm{\Gamma}\\
\end{aligned}
\end{equation}
where the former is the DO nonlocal traction boundary condition (TBC) and the latter is the displacement boundary condition (DBC). While, on the surface, the above derivation appears equivalent to the classical elastodynamic formulation, there are important details to be considered. Notably, the R-RL type DO operator within the stress field is self-adjoint in nature and this specific behavior, along with the underlying symmetry in $\bm{C}$ and $\bm{\kappa}$, simplifies significantly the variational calculations in Eq.~(\ref{ref: EOM_simplification}). A detailed discussion on self-adjointness, symmetry, and the variational calculations can be found in \S S3, \S S4, and \S S5 of the SM. Moreover, it is noteworthy that the use of DO fractional constitutive relation (see Eq.~(\ref{eq: DO constitutive relation})) does not affect the general form of the nonlocal governing equations and the corresponding boundary conditions.

%%%%%%%%%%%%%%%%%%%%%%%%%%%%%%%%%%%%%%%%%%%%%%%%%%%%%%%%%%%%%%%%
\section{Physical interpretation of DO nonlocal elasticity theory}
\label{sec: 4}
The previous sections have presented a detailed mathematical formulation of the DO-NET. To be able to apply this formulation to the solution of practical problems, it is fundamental to establish a connection between the mathematical structure and the physical behavior of the medium. To address this aspect, we employ the MSLM to develop a mechanical structure that is physically equivalent to the DO-NET and that can be used to derive some high level understanding of the connections between discrete scales (typical of microstructural or even molecular models) and continuum scales. The MSLM has found many previous applications in multiscale nonlocal elasticity problems~\cite{polyzos2012derivation,carpinteri2014nonlocal}. By interpreting nonlocal interactions in continuous media as the continuum limit of spring forces between discrete lattice points, MSLM can be used to establish more direct connections between the nonlocal theoretic formulation and the physical characteristics of the models. These models, often used to represent the medium at microstructural level, provide a direct approach to interpret the multiscale nature captured by DO-NET (see Fig.(\ref{fig: DO_system})). In this regard, not only the MSLM can be used to model DO nonlocal elasticity problems, but it can be leveraged to unravel the physical implications of DO-NET at the micro scales.

In the following, we illustrate the approach by taking a one-dimensional nonlocal rod as a sample problem. The rod is assumed to have a non-uniform distribution of nonlocality along the $y$ (thickness) direction.
The rod will be modeled based on the MSLM and compared with the continuum DO formulation. The equivalence of these two models will be shown both in terms of the linear momentum and of the elastic energy. The resulting equivalent MSLM model will allow drawing important conclusions about the DO-NET formulation.
The nonlocal rod example was chosen to maintain a fairly straightforward mathematical formulation, hence facilitating the understanding of the DO-NET and of the correspondences with the MSLM model. At the same time, the results are general and provide considerations directly applicable to a three dimensional continuum.

\subsection{Mass spring lattice model (MSLM) of a nonlocal rod}\label{ssec: 4.1}
\begin{figure}[ht]
    \centering
    \includegraphics[width=0.9\linewidth]{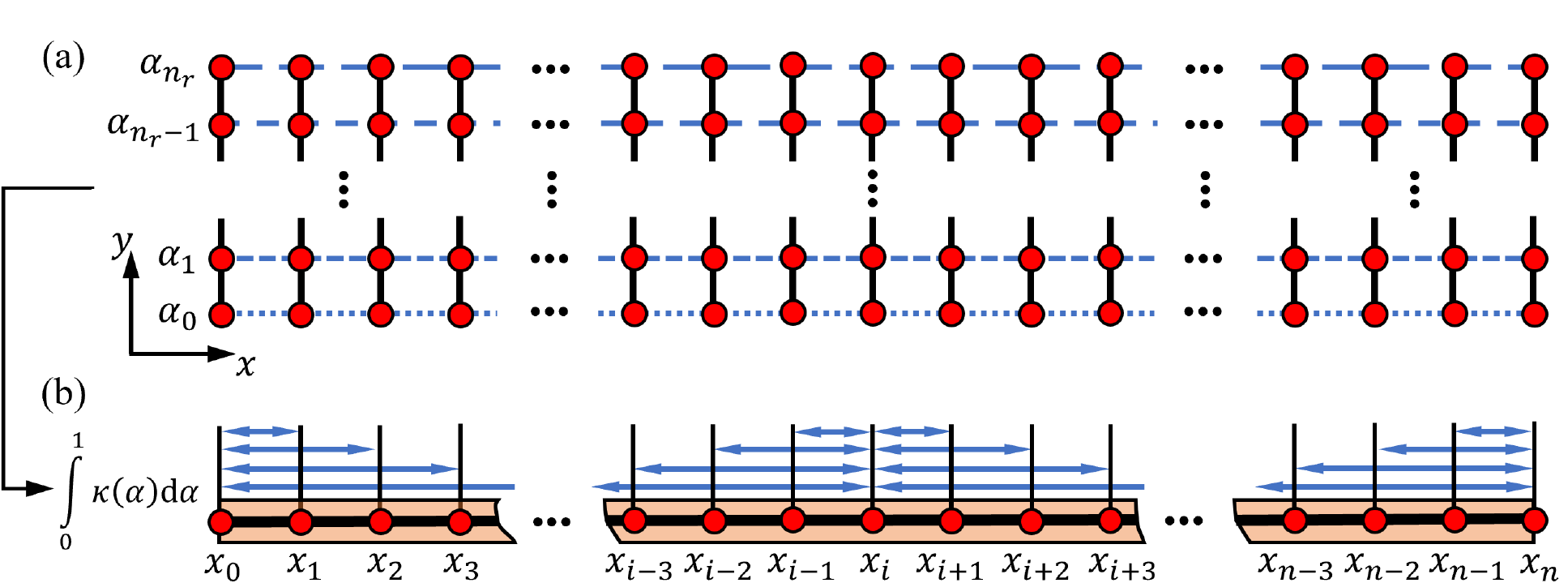}
    \caption{Illustration of the MSLM equivalent to a 1D finite DO nonlocal rod. (a) Finite 2D MSLM with transversely distributed order of nonlocality $\alpha_r$, $r=0,1,...,n_r-1,n_r$. Different dashed line styles represent different levels of nonlocality in each layer. (b) The equivalent 1D MSLM (red circles connected by solid black line) overlapped with the 1D nonlocal rod with a DO operator $\int_{0}^{1}\kappa(\alpha)\mathrm{d}\alpha$ (orange rectangular box). By combining all the parallel nonlocal interactions, the $y$ direction in the 2D lattice system can be reduced so to obtain an equivalent 1D MLSM. The 1D MSLM consists of uniformly distributed $n+1$ mass points (red circles) and DO nonlocal interactions modeled by elastic springs (shown as double arrow blue lines) on the $x$ direction. The corresponding 1D rod governed by DO-NET (shown as a rectangle in orange) has length $L=n\Delta$, where $\Delta$ is the distance between two nearest lattice points (or mesh size) in the equivalent 1D MSLM. Specifically, for a given mass point $x_i$ in MSLM, springs are placed on both its left and right sides. Unlike the infinite lattice system in Fig.~(\ref{fig: DO_system}), springs are placed only on either the right or left side at the boundary points $x_0$ and $x_n$, respectively.}
    \label{fig: MSLM}
\end{figure}

Consider a rod having a non-uniform distribution of nonlocality along the $y$ (thickness) direction. The system can be seen as a layered rod where each layer exhibits nonlocal behavior in the $x$ direction and where the degree of nonlocality changes across different layers (but it is constant along the same layer). This system can be modeled via MSLM by modifying the 2D unbounded lattice system in Fig.~(\ref{fig: DO_system}) to represent a finite domain (see Fig.~(\ref{fig: MSLM})). This lattice can also be considered as an equivalent (simplified) microscale representation of the nonlocal rod and it will be helpful to provide a practical physical interpretation of DO-NET. The development of the MSLM can be broadly divided into the following steps:
\begin{enumerate}[label={[S\arabic*]}]
    \item The nonlocal rod, whose multiple scale nature originates from the non-uniform (transverse) distribution of nonlocality, is translated into a finite 2D lattice system consisting of nonlocal chains stacked along the transverse direction. The nonlocal chains stacked in the transverse direction have different degree of nonlocality, similar to the unbounded 2D lattice system introduced in \S\ref{sec: 2} (see Fig.~(\ref{fig: DO_system})). 
    
    \item Recall from the discussion in \S\ref{sec: 2} that the multiscale nonlocal effect, resulting from the transverse variation of properties in the 2D lattice system, can be captured within a DO derivative defined on the axial (1D) direction. This is evident from Eq.~(\ref{eq: DO_hom_tot_F_cont_step3}), where the effect of the underlying multiple scales (stacked in the transverse direction) is captured within the strength function of the DO derivative and the nonlocality in each scale is captured via the CO derivative (within the DO derivative). In conclusion, we note that the DO derivative enables to reduce the 2D description following the presence of multiple scales to a simpler (1D) description.
    
    \item Following the above discussion, the 2D finite lattice system describing the nonlocal rod (in [S1]) can be \textit{effectively reduced} to a 1D lattice system where the long-range connections are captured via DO derivatives. In this 1D lattice system, the information of the multiple scales underlying the nonlocal rod, will be captured in the strength function $\kappa(\alpha)$. In more specific terms, the summation of all the spring forces (from the different material scales) at a given point in the 2D lattice system should be captured effectively in a 1D DO form of the stress-strain constitutive relation (see Eq.~(S36)). This reduction scheme is also illustrated in Fig.~(\ref{fig: MSLM}). In the remaining study, we will refer to the 1D lattice system as the 1D DO MSLM (or simply, MSLM) of the nonlocal rod.
    
    \item The derivation of the MSLM model reduces to estimating the different spring stiffness such that the (1D) DO nonlocal equation of motion (EOM) and boundary conditions (BC) are obtained from the MSLM upon continualization. The details of the 1D DO-NET, obtained by assuming $\bm{u}(\bm{x})\equiv u(x)$ (that is, only axial deformation occurs) in Eq.~(\ref{eq: EOM}), is provided in SM~\S{6}. The equivalence between the two formulations (that are DO-NET and MSLM) will be assessed both by evaluating the linear momentum and the deformation energy. Note that the steps through [S1]-[S3] are self-contained and complete. In the following we will focus specifically on [S4], where we develop the equivalent spring stiffness for the 1D DO MSLM.
\end{enumerate}

\subsubsection{Linear momentum equivalence}\label{sssec: 4.1.1}
Consider the simplified 1D DO MSLM (see Fig.~(\ref{fig: MSLM}-b)) obtained by reducing the order of the finite 2D lattice (see Fig.~(\ref{fig: MSLM}-a)). By using Newton's second law, the EOM of a given particle $i$ of the lattice can be written as:
\begin{equation}\label{eq: 1D EOM MSLM}
    \sum_{j=0,j \neq i}^{n}k_{ij}(u_j-u_i)+F^{(E)}_i=m\Ddot{u}_i ~~ \forall ~i \in [1,n-1]
\end{equation}
where $k_{ij}$ denotes the spring stiffness between point $i$ and $j$ equivalent to the set of parallel springs in the $y$ direction, $n$ is the total number of particles, and $F^{(E)}_i$ denotes the external force acting on the particle $i$. By taking the continuum limit Eq.~(\ref{eq: 1D EOM MSLM}) can be expressed as:
\begin{equation}
\label{eq: 1D EOM MSLM continuum limit}
   \lim_{\Delta \to 0}\frac{1}{\Delta}\left[\sum_{j=0,j \neq i}^{n}k_{ij}(u_j-u_i)+F^{(E)}_i\right]
   =\lim_{\Delta \to 0}\frac{m}{\Delta}\Ddot{u}_i~\Rightarrow~ \underbrace{\lim_{\Delta \to 0}\frac{1}{\Delta}\left[\sum_{j=0,j \neq i}^{n}k_{ij}(u_j-u_i)\right]}_{\phi}+f^{(E)}(x_i)=\tilde{\rho}\Ddot{u}_i
\end{equation}
where $\Delta=L/n$ is the distance between two nearest points, $L$ is the length of the lattice in the $x$-direction, and $f^{(E)}(\cdot)$ is the force density function with dimensions of force per unit length $[\textrm{N}/\textrm{m}]$. As evident from the above expression $\tilde{\rho}$ ($= \lim_{\Delta \to 0}({m}/{\Delta})$) denotes the mass per unit length of the lattice. Note that, the term $\phi$ indicated in the above expression should converge to the divergence of stress in the continuum description. Hence, in order to enforce the equivalence between the DO-NET and 1D MSLM, it is required that:
\begin{equation}\label{eq: spring force continuum limit_DO}
    \lim_{\Delta \to 0}\frac{1}{\Delta}\left[\sum_{j=0,j \neq i}^{n}k_{ij}(u_j-u_i)\right] = EA\int_{0}^{1}\kappa(\alpha)\left({D}\prescript{R-M}{0}{}{\bm{D}}_{L}^{\alpha}u(x)\right)\textrm{d}\alpha = 
    A\left[D\sigma\right] \equiv A\left[\nabla{\cdot}\sigma\right]
\end{equation}
where $E$ and $A$ denote the Young's modulus and cross-sectional area of the 1D nonlocal solid, respectively. In Eq.~(\ref{eq: spring force continuum limit_DO}) above, we have used Eq.~(S39) to express the divergence of the stress in the 1D governing equation (see SM~\S{S6}). As evident from the above equation, the development of the MSLM equivalent to the continuum DO formulation reduces to the evaluation of the terms $k_{ij}$, as highlighted previously in \S\ref{ssec: 4.1}. It appears that a straightforward and, in fact, natural approach to determining the $k_{ij}$ consists in a numerical discretization of the DO term $D\sigma(x_i)$. Note that Eq.~(\ref{eq: spring force continuum limit_DO}) holds because in the 1D case the divergence of the DO stress $\nabla{\cdot}\sigma$ is equivalent to the first order derivative of the DO stress $D\sigma$.

Broadly speaking, the numerical approximation of DO operators consists of two key steps: 1) approximation of the integral operator $\int_0^1(\cdot)\textrm{d}\alpha$, and 2) approximation of the constant fractional operator ${D}\prescript{R-M}{a}{}{\bm{D}}_b^{\alpha}(\cdot)$~\cite{ding2021applications}. Note that step 1 requires a discretization of the interval of the order $[0,1]$, and step 2 requires a discretization over the spatial domain $[0,L]$. The discretization of the order interval in step 1 yields \cite{ding2021applications}:
\begin{equation}\label{eq: spring force continuum limit step1}
    \lim_{\Delta \to 0}\frac{1}{\Delta}\left[\sum_{j=0,j \neq i}^{n}k_{ij}(u_j-u_i)\right]= EA \sum_{r=0}^{n_\alpha}w_r\kappa(\alpha_r)\Delta_\alpha \Big[ \underbrace{{D}\prescript{R-M}{0}{}{\bm{D}}_{L}^{\alpha_r} u(x_i)}_{\text{Step 2}} \Big]
\end{equation}
where the distributed order interval $[0,1]$ was divided uniformly into $n_\alpha$ increments with size $\Delta_\alpha = 1/n_\alpha$. $w_r$ is a set of numerical integration weights, which is determined from the numerical technique followed to approximate the integral \cite{ding2021applications}. The expression in square brackets is the constant order fractional operator that must be approximated in step 2. In order to differentiate the parameters for spatial discretization (over a mesh of points $x_i$ with $i=0,...,n$, see Fig.~(\ref{fig: MSLM})), we also use $\alpha_r$ with $r=0,...,n_\alpha$ to denote distributed order discretization in step 1 approximation. In step 2, each fractional-order derivative (with order $\alpha_r$, $r=0,...,n_\alpha$) in Eq.~(\ref{eq: spring force continuum limit step1}) is approximated independently such that:
\begin{equation}\label{eq: spring force continuum limit}
    \lim_{\Delta \to 0}\frac{1}{\Delta}\left[\sum_{j=0,j \neq i}^{n}k^{\alpha_r}_{ij}(u_j-u_i)\right]={EA}\left[{D}\prescript{R-M}{0}{}{\bm{D}}_{L}^{\alpha_r} u(x_i)\right]
\end{equation}
where $k^{\alpha_r}_{ij}$ denotes the nonlocal spring stiffness resulting from an infinitesimal order element within the order interval $[0,1]$. Before proceeding further with the numerical approximation, we note from a physical perspective that, $k_{ij}^{\alpha_r}$ can be regarded as the strength of the nonlocal interactions in a single layer in the 2D lattice, and $k_{ij}$ accounts for the total parallel nonlocal interactions of all layers along y-direction. In this regard, the total nonlocal spring stiffness between the points $i$ and $j$ can be expressed by combining Eqs.~(\ref{eq: spring force continuum limit step1},\ref{eq: spring force continuum limit}) as:
\begin{equation}
    k_{ij} = \sum_{r=0}^{n_{\alpha}}w_r\kappa(\alpha_r) k^{\alpha_r}_{ij}\Delta_{\alpha}
\end{equation}
The above expression further reinforces the role of DO operators to account for nonlocal effects characterized by long-range interactions acting in parallel. This observation is consistent with the previous multiscale configuration in \S\ref{sec: 2} and also further substantiates the equivalence between the 1D DO rod and the MSLM.

Note that the approximation in Eq.~(\ref{eq: spring force continuum limit}) applies to any order $\alpha_r \in (0,1)$. Hence, in the interest of a more compact notation, in the following derivation we will drop the subscript $r$ and simply denote ${\alpha_r}$ as ${\alpha}$ and $k_{ij}^{\alpha_r}$ as $k_{ij}^{\alpha}$. The detailed expression for the differ-integral operator $\left({D}\prescript{R-M}{0}{}{\bm{D}}_{L}^{\alpha_r} u(x_i)\right)$ can be found in Eqs.~(S39,S40) of \S{S6}. Further, the details of the numerical approximation for $ {D}\prescript{R-M}{0}{}{\bm{D}}_{L}^{\alpha_r} u(x_i)$  can be found in \S{S7} of the SM. Thanks to the R-M definition, the numerical approximation of the aforementioned differ-integral operator contains the relative displacement terms $(u(x_j)-u(x_i) \equiv u_j - u_i)$ (see Eq.~(S45) of SM \S{S7}). Indeed, recalling the configuration of MSLM in Eq.~(\ref{eq: 1D EOM MSLM}) and the requirement of model equivalence in Eq.~(\ref{eq: spring force continuum limit}) for a given point $x_i$, the stiffness $k_{ij}^{\alpha}$ of the nonlocal spring between $x_i$ and $x_j$ can be read off from the coefficient of the term $(u(x_j)-u(x_i))$ (of Eq.~(S45) in SM \S{S7}) as:
\begin{equation}\label{eq: k_ij}
    k_{ij}^{\alpha}=\frac{{EA}\Delta}{2\Gamma(1-\alpha)}
    \begin{cases}
    \alpha(x_i-x_j)^{-(1+\alpha)}+\alpha(1+\alpha)(x_i-x_j)^{-(2+\alpha)}\Delta & j=0\\
    \alpha(1+\alpha)(x_i-x_j)^{-(2+\alpha)}\Delta & 0<j<i-1\\
    \frac{\alpha(1+\alpha)}{1-\alpha}\Delta^{-(1+\alpha)} & j=i-1\\
    \frac{\alpha(1+\alpha)}{1-\alpha}\Delta^{-(1+\alpha)} & j=i+1\\
    \alpha(1+\alpha)(x_j-x_i)^{-(2+\alpha)}\Delta & i+1<j<n\\
    \alpha(x_j-x_i)^{-(1+\alpha)}+\alpha(1+\alpha)(x_j-x_i)^{-(2+\alpha)}\Delta & j=n\\
    \end{cases} {~~~~ \forall ~i \in [1,n-1]}
\end{equation}
Further, the stiffness of the springs connecting the boundaries to their adjacent points are obtained as:
\begin{equation}\label{eq: k_ij limit case}
\begin{aligned}
    k^\alpha_{01}&=\frac{EA\Delta}{2\Gamma(1-\alpha)}\left[\alpha(x_1-x_0)^{-(1+\alpha)}+\frac{\alpha(1+\alpha)}{1-\alpha}(x_1-x_0)^{-(1+\alpha)}\right] \equiv \frac{EA}{\Gamma(1-\alpha)}\left[\frac{\alpha}{1-\alpha} \Delta^{-\alpha}\right]\\
    k^\alpha_{n-1,n}&=\frac{EA\Delta}{2\Gamma(1-\alpha)}\left[\alpha(x_n-x_{n-1})^{-(1+\alpha)}+\frac{\alpha(1+\alpha)}{1-\alpha}(x_n-x_{n-1})^{-(1+\alpha)}\right] \equiv \frac{EA}{\Gamma(1-\alpha)}\left[\frac{\alpha}{1-\alpha} \Delta^{-\alpha}\right]\\
\end{aligned}
\end{equation}
where $k^\alpha_{n-1,n}$\footnote{The comma in the subscript of $k^\alpha_{n-1,n}$ is used to separate $n-1$ and $n$ to avoid confusion in the notation, and does not indicate any derivative (as usually done in indicial notation).} denotes the spring stiffness between $n-1$ and $n$. A detailed treatment of above two cases (and the corresponding spring constants) can be found in SM~\S{S8}. This completes the derivation of the lattice equivalent to the continuum EOM for all body points within the 1D solid except the only remaining term $k_{0n}^{\alpha}$ that connects two boundary points $x_0$ and $x_n$. 

Note that to further guarantee the equivalence between the MSLM and the DO-NET, the $k_{0n}^{\alpha}$ should be determined so that the total spring forces acting at the boundary points $x_0$ and $x_n$ are equivalent to continuum boundary conditions defined in the DO-NET. For this purpose, we adapt the strategy followed in deriving the stiffness of the bulk nonlocal connections. More specifically, we aim to obtain: 1) the expression of the force developed at the two boundary points $x_0$ and $x_n$ from the nonlocal spring interactions in the MSLM, and 2) the discrete (numerical) approximation of the continuum boundary force in DO-NET, in terms of the relative displacements between the different points. Finally by comparing directly the coefficient of $u_n - u_0$, we read off the nonlocal stiffness $k^\alpha_{0n}$.

We consider the derivation at left boundary point $x_0$. The total force on $x_0$ due to the nonlocal connections denoted by $F_0^{M}$, is obtained by using balance of forces as:
\begin{equation}\label{eq: F_0_step1}
F_0^{M} = {\sum_{j=1}^{n-1}k^\alpha_{0j}(u_j-u_0)} + k^{\alpha}_{0n}(u_n - u_0)
\end{equation}
where $k^\alpha_{0n}$ is to be determined. The superscript $'M'$ in $F_0^M$ indicates that the same was obtained from the MSLM. Now, by using the expressions for the stiffness of the different nonlocal springs from Eq.~(\ref{eq: k_ij}) we obtain:
\begin{equation}\label{eq: Explicit F_0_step1}
\begin{split}
F_0^{M} =\frac{{EA}\Delta}{2\Gamma(1-\alpha)}\left[\sum_{i=1}^{n-1}\alpha\frac{u_i-u_0}{(x_i-x_0)^{1+\alpha}}+\Delta\sum_{i=2}^{n-1}\alpha(1+\alpha)\frac{u_i-u_0}{(x_i-x_0)^{2+\alpha}}+\frac{\alpha(1+\alpha)}{1-\alpha}\frac{u_1-u_0}{\Delta^{1+\alpha}}\right]\\ + k^\alpha_{0n}(u_n - u_0)
\end{split}
\end{equation}
It appears that, in order to derive an expression for $k^{\alpha}_{0n}$, we must derive an expression for $F_0^{M}$.

To guarantee the boundary equivalence between the MSLM and the DO-NET, the discrete MSLM-based formulation of the total force at the boundary in Eq.~(\ref{eq: F_0_step1}) (that is, $F_0^{M}$) should be equivalent to the tractions defined in 1D DO-NET (see Eq.~(S36) of \S{S6}) upon continualization. Recall that, for the classical (local) MSLM with connections only between nearest-neighbor lattice points~\cite{thomas2006lattice}, the spring force $F_0^{M(l)}$ acting at the boundary point $x_0$ can be expressed using Taylor's approximation as:
\begin{equation}\label{eq: F_0 integer}
    F_{0}^{M (l)}=k_{01}(u(x_1)-u(x_0)) \approx k_{01}{\Delta}\left(D u(x_0)+\frac{\Delta}{2}D^2 u(x_0)\right)
\end{equation}
$k_{01}$ denotes the spring stiffness between $x_0$ and $x_1$ for local the MSLM. The continuum limit for the above expression is obtained as:
\begin{equation}\label{eq: F_0 integer_step1}
    F_{0}^{C (l)}= \lim_{\Delta{\to}0} k_{01}{\Delta}\left(D u(x_0)+\frac{\Delta}{2}D^2 u(x_0)\right) = E A \left[ D u(x)\big|_{x=x_0}+ \lim_{\Delta{\to}0}\left(\frac{\Delta}{2} D^2u(x)\right)\bigg|_{x=x_0} \right]
\end{equation}
where we used $\lim_{\Delta{\to}0} k_{01}{\Delta} = EA$. Note that the two differential terms $Du(x)$ and $D^2u(x)$ correspond to the boundary conditions and the governing equation at $x=x_0$, respectively~\cite{thomas2006lattice}. Inspired by the local MSLM approach shown in Eq.~(\ref{eq: F_0 integer}), it can be proved that total forces acting at $x=x_0$ in DO nonlocal MSLM can be formulated as:
\begin{equation}
    F_0^C = \lim_{\Delta{\to}0}F_0^M = \lim_{\Delta{\to}0}EA\left[\left(\frac{1}{2}\prescript{M}{x_0-\Delta}{}{\bm{D}}_{x_0}^{\alpha}u(x)-\frac{1}{2}\prescript{M}{x_0}{}{\bm{D}}_{x_n}^{\alpha}u(x)\right)\bigg|_{x=x_0}+\left(-\frac{\Delta}{2}{D}\prescript{M}{x_0}{}{\bm{D}}_{x_n}^{\alpha}u(x)\right)\bigg|_{x=x_0}\right]
\end{equation}
with the previously non-determined nonlocal stiffness $k_{0n}^{\alpha}$ given as:
\begin{equation}
\label{eq: boundary_stiffness}
    k^\alpha_{0n}=k^\alpha_{n0}=\frac{EA}{2\Gamma(1-\alpha)}\left[ (x_n-x_0)^{-\alpha} + \alpha\Delta (x_n-x_0)^{-(1+\alpha)} + \alpha(1+\alpha)\Delta^2 (x_n-x_0)^{-(2+\alpha)} \right]
\end{equation}
Detailed derivation of $k^\alpha_{0n}$ is provided in SM~\S{S9}. The description above completes the derivation of all spring stiffness terms within the nonlocal MSLM and establishes the equivalence between the MSLM and the DO-NET for all the lattice points including boundary and body points. More importantly, with the continuum expression obtained for nonlocal MSLM, the equivalence of the traction boundary conditions (TBC) between MSLM and DO-NET can be further shown, hence strengthening the ability of DO-NET to properly represent the physical system at the micro-scales. Detailed proof of the TBC equivalence is provided in SM~\S{S10}.

\subsubsection{Energy equivalence}\label{sssec: 4.1.2}
In this section, we further substantiate the equivalence between the continuum limit of the MSLM and DO-NET on the basis of energy arguments. In addition to establishing the equivalence from an energy perspective, we will also show how the energy approach helps isolating and characterizing surface effects due to nonlocality \cite{li2020contribution}. Note that, although we started from a linear momentum equivalence approach in order to derive the MSLM and then followed up demonstrating the energy equivalence, the opposite path (that is, starting from an energy equivalence to derive the MSLM and then demonstrating the force-equivalence) will yield identical results. We merely note that this latter approach is more useful to derive higher dimensional MSLM since they enable the application of variational-principles to derive governing equations in a straightforward fashion.

Similar to \S\ref{ssec: 3.2}, we start by deriving the explicit expressions for the total potential energy obtained via both the DO-NET and the MSLM. For the DO-NET, following Eq.~(\ref{eq: Explicit potential energy}), the total potential energy for the 1D DO nonlocal rod can be obtained as:
\begin{equation}\label{eq: Pi^C1}
    \Pi^{C_1}=\int_{0}^{L} \underbrace{\frac{1}{2}EA\left[Du(x)\mathcal{D}^{\alpha}u(x)\right]}_{\mathbb{U}^{C_1}(x)}\textrm{d}x
\end{equation}
$\mathbb{U}^{C_1}(x)$ denotes the potential energy density function. In the above equation, we used the superscript $'{C_1}'$ to denote the total potential energy derived at continuum level. Similarly, for the discrete MLSM based on the formulation of force equivalence in \S\ref{sssec: 4.1.1}, the total energy stored in the MLSM's springs is given by:
\begin{equation}\label{eq: Pi^M}
    \Pi^M=\sum_{i=0}^{n}\mathbb{U}_i^M=\sum_{i=0}^{n}\left[\frac{1}{4}\sum_{j=0,j \neq i}^{n}\left(\int_{0}^{1}\kappa(\alpha)k_{ij}(\alpha)\textrm{d}\alpha\right)(u_j-u_i)^2\right]
\end{equation}
where $\mathbb{U}_i^M$ is the elastic energy stored in all the springs (local as well as nonlocal) connecting to a given point $i$. The superscript $'M'$ in the above equation is used to distinguish the energy obtained via the MSLM from the DO continuum theory. Note that an additional multiplicative factor $1/2$ is added in the potential energy of the springs in the above equation in order to equally distribute the spring potential energy between the two points connected via a given spring. Using Eqs.~(\ref{eq: Pi^C1},\ref{eq: Pi^M}), we will establish an exact match between the potential energies obtained via the DO continuum formulation and the MSLM.

By expanding the relative displacement term $(u_j-u_i)^2$ using binomial theorem, the $\mathbb{U}_i^M$ term in Eq.~(\ref{eq: Pi^M}) can be expressed in a straightforward manner as:
\begin{equation}\label{eq: U_i^M}
    \mathbb{U}_i^{M}=\frac{1}{4}\sum_{j=0,j \neq i}^{n}\left(\int_{0}^{1}\kappa(\alpha)k_{ij}(\alpha)\textrm{d}\alpha\right)\left(u_j^2-u_i^2\right)-\frac{1}{2}u_i\sum_{j=0,j \neq i}^{n}\left(\int_{0}^{1}\kappa(\alpha)k_{ij}(\alpha)\textrm{d}\alpha\right)\left(u_j-u_i\right)\\
\end{equation}
Using the above simplification, we transferred the algebraic expression $(u_j-u_i)^2$ in $\mathbb{U}_i^M$ into $u_j^2-u_i^2$ and $u_j-u_i$, which enables a simplification of the resulting expressions in terms of the spring forces in Eq.~(\ref{eq: 1D EOM MSLM}). Now, by multiplying and dividing the right-hand side of the above expression by the inter-particle spacing $\Delta$, and then taking the continuum limit, we obtain:
\begin{equation}\label{eq: continuum limit U_i^M}
    \lim_{\Delta \to 0}\mathbb{U}_i^{M}=\lim_{\Delta \to 0}{\Delta} \underbrace{EA\left(\frac{1}{4}D\mathcal{D}^{\alpha}u^2(x_i)-\frac{1}{2}u(x_i)D\mathcal{D}^{\alpha}u(x_i)\right)}_{\mathbb{U}^{C_2}(x_i)}
    =\lim_{\Delta \to 0}{\Delta}\mathbb{U}^{C_2}(x_i)
\end{equation}
for body points. Following the same procedure, we also obtain the following expressions:
\begin{equation}\label{eq: continuum limit U_0^M and U_n^M}
\begin{aligned}
    \lim_{\Delta \to 0}\mathbb{U}_0^M&=\lim_{\Delta \to 0}{\Delta}EA\left(\frac{1}{4}D\mathcal{D}^{\alpha}u^2(x_0)-\frac{1}{2}u(x_0)D\mathcal{D}^{\alpha}u(x_0)\right)+EA\left(\frac{1}{4}\mathcal{D}^{\alpha}u^2(x_0)-\frac{1}{2}{u(x_0)}\mathcal{D}^{\alpha}u(x_0)\right)\\
    &=\lim_{\Delta \to 0}{\Delta}\mathbb{U}^{C_2}(x_0)+\mathbb{U}^{b}(x_0)\\
    \lim_{\Delta \to 0}\mathbb{U}_n^M&=\lim_{\Delta \to 0}{\Delta}EA\left(\frac{1}{4}D\mathcal{D}^{\alpha}u^2(x_n)-\frac{1}{2}u(x_n)D\mathcal{D}^{\alpha}u(x_n)\right)-EA\left(\frac{1}{4}\mathcal{D}^{\alpha}u^2(x_n)-\frac{1}{2}{u(x_n)}\mathcal{D}^{\alpha}u(x_n)\right)\\
    &=\lim_{\Delta \to 0}{\Delta}\mathbb{U}^{C_2}(x_n)-\mathbb{U}^{b}(x_n)\\
\end{aligned}
\end{equation}
for the two boundary points. Comparing Eq.~(\ref{eq: continuum limit U_i^M}) and Eq.~(\ref{eq: continuum limit U_0^M and U_n^M}), it can be found that total energy at the boundary points contain two terms $\mathbb{U}^{b}(x_0)$ and $\mathbb{U}^{b}(x_n)$ that are independent of the discretization, and a common term $\mathbb{U}^{C_2}(x_i)$ which matches exactly the expression of the potential energy at the internal points (see Eq.~(\ref{eq: continuum limit U_i^M})). Consequently, from Eqs.~(\ref{eq: continuum limit U_i^M},\ref{eq: continuum limit U_0^M and U_n^M}) we obtain that:
\begin{equation}\label{eq: U^C2}
    \mathbb{U}^{C_2}(x)=\lim_{\Delta \to 0}\frac{1}{\Delta}\mathbb{U}^{M}(x)=EA\left(\frac{1}{4}D\mathcal{D}^{\alpha}u^2(x)-\frac{1}{2}u(x)D\mathcal{D}^{\alpha}u(x)\right)
\end{equation}
such that the total elastic energy stored in MSLM can be expressed as:
\begin{equation}\label{eq: Pi^M and Pi^C1}
\begin{aligned}
    \lim_{\Delta{\to}0}\Pi^{M}&=\Pi^{C_2}=\int_{0}^{L}\mathbb{U}^{C_2}(x)\textrm{d}x+\mathbb{U}^{b}(0)-\mathbb{U}^{b}(L)\\
    &=\underbrace{EA\left(\frac{1}{4}\mathcal{D}^{\alpha}u^2(x)-{\frac{1}{2}u(x)}\mathcal{D}^{\alpha}u(x)\right)\bigg|_{0}^{L}}_{\Pi^M_1}+\underbrace{\frac{1}{2}\int_{0}^{L}EA\left(Du(x)\mathcal{D}^{\alpha}u(x)\right)\textrm{d}x}_{\Pi^M_2}+\underbrace{\mathbb{U}^{b}(0)-\mathbb{U}^{b}(L)}_{\Pi^M_3}\\
    &=\Pi^{C_1} \\
\end{aligned}
\end{equation}
In the above equation, the terms $\Pi^M_1$ and $\Pi^M_2$ are obtained via integration by parts of the expression for $\mathbb{U}^{C_2}(x)$ from Eq.~(\ref{eq: continuum limit U_i^M}). Note that the term $\Pi^M_1$ exactly cancels the term $\Pi^M_3$. This establishes the equivalence of $\Pi^M$, $\Pi^{C_1}$, and $\Pi^{C_2}$. Note that while $\Pi^{C_1}$ and $\Pi^{C_2}$ represent the same physical quantity, that is the potential energy of the continuum, there is a difference in the procedure adopted to obtain the two expressions. While $\Pi^{C_1}$ was obtained directly from the continuum expression of the potential energy density $\mathbb{U}^{C_1}$, $\Pi^{C_2}$ was obtained by starting from the discrete expressions of the potential energy density $\mathbb{U}_i^M$. Hence, we chose to denote them separately. Further, note that for $\kappa(\alpha)=\delta(1)$, the DO nonlocal operator becomes a local integer operator such that both $\mathbb{U}^{b}(0)=\mathbb{U}^{b}(L)=0$. This indicates that the boundary energy terms in $\Pi^M_3$ arise primarily due to nonlocality and can be interpreted as a surface effect due to the truncation of the nonlocal interactions at the boundary. This latter observation also coincides with investigations conducted via classical approaches to nonlocal elasticity \cite{wang2011vibration,li2020contribution}. Accounting for surface effects is critical in several applications at the nano- and micro-scales \cite{wang2011vibration,narendar2012study,hosseini2018nonlocal,li2020contribution}.

%%%%%%%%%%%%%%%%%%%%%%%%%%%%%%%%%%%%%%%%%%%%%%%%%%%%%%%%%%%%%%%%
\section{Numerical examples}
\label{sec: 5}
\begin{figure}[ht]
    \centering
    \includegraphics[width=0.8\linewidth]{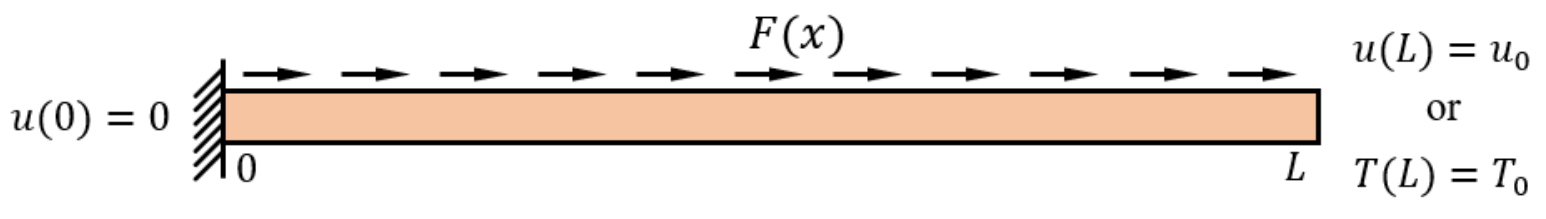}
    \caption{Schematic of a 1D nonlocal rod under a distributed axial load $F(x)$. This structure is used as benchmark problem to evaluate the performance of the DO formulation. The left boundary is fixed while the right boundary can be subject to either a prescribed displacement $u_0$ or traction $T_0$.}
    \label{fig: 1D rod}
\end{figure}

In this section, we show via numerical means the equivalence between the MSLM and the proposed DO-NET. We consider the continuum limit of the mass-spring lattice presented in \S\ref{sec: 4} (a 1D nonlocal rod) in Fig.~(\ref{fig: 1D rod}). For the sake of simplicity and without loss of generality, we assume that the length and area-normalized Young's modulus of 1D rod are $L=1\textrm{m}$ and $E=1\textrm{Pa}/\textrm{m}^2$, respectively. 

Before proceeding to present the numerical results, we briefly discuss the procedure adopted to obtain the numerical solutions of the MSLM and the DO-NET. The MSLM is already in a discretized form that is amenable to a straightforward numerical implementation; recall the set of algebraic equations in \S\ref{ssec: 4.1}. 
The solution for the DO-NET requires some more discussion. 
Recall that the approximation of DO operators is divided into two steps. For the first step, involving the approximation of integral operator $\int_{0}^{1}(\cdot)\mathrm{d}\alpha$, we use the trapezoidal scheme~\cite{patnaik2020application,ding2021applications}. Using the trapezoidal scheme, the fractional stress $\sigma$ in Eq.~(S24) can be approximated as a multi-term (discrete) DO fractional derivative as:
\begin{equation}\label{eq: multi-term sigma}
\begin{aligned}
    \sigma(x) \approx \sum_{r=0}^{n_\alpha-1}\frac{\Delta_\alpha}{2}\left[\frac{\kappa(\alpha_{r})}{2}\left(\prescript{C}{0}{}{\bm{D}}_x^{\alpha_{r}}u(x)-\prescript{C}{x}{}{\bm{D}}_{L}^{\alpha_{r}}u(x)\right)+\frac{\kappa(\alpha_{r+1})}{2}\left(\prescript{C}{0}{}{\bm{D}}_x^{\alpha_{r+1}}u(x)-\prescript{C}{x}{}{\bm{D}}_{L}^{\alpha_{r+1}}u(x)\right)\right]
\end{aligned}
\end{equation}
which is composed of $n_\alpha$ number of CO derivatives. In the above approximation, the discretization for the order integral is identical to Eq.~(\ref{eq: spring force continuum limit step1}). Further, the same discretization is also utilized to evaluate the derivative of the stress field, $D\sigma(x)$. For step two, involving the approximation of the CO fractional derivative with order $\alpha_{i}$, we use the rectangular rule outlined previously in SM~\S{S7} (see Eq.~(S42-S44)). While it is possible to adopt other techniques, this choice ensures that the same level of numerical approximation is used in both the DO-NET and the MSLM models (recall that the rectangular rule was used in the process of deriving MSLM in \S\ref{sec: 4}).

\subsection{Numerical results}\label{ssec: 5.1}
\begin{figure}[ht!]
    \centering
    \includegraphics[width=\textwidth]{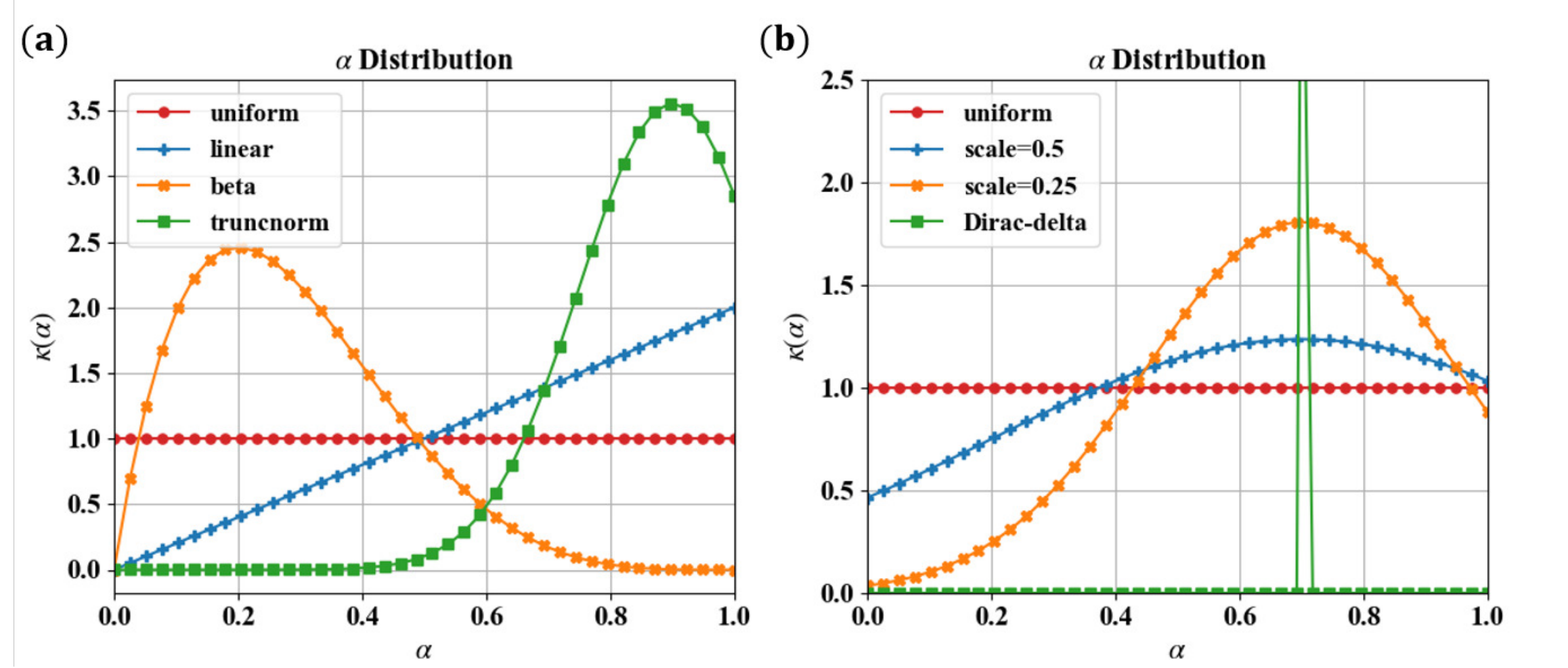}
    \caption{Strength-functions $\kappa(\alpha)$ for two test cases. (a) shows test case 1 containing the strength-functions $\kappa(\alpha)$ of four classical continuous distribution with finite support on $[0,1]$. Specifically, the uniform, linear, beta (with $a=2$ and $b=5$), and truncated normal (with $\textrm{loc}=0.9$ and $\textrm{scale}=0.15$) distributions. (b) shows test case 2 containing the strength-functions $\kappa(\alpha)$ of four evolutionary continuous distributions. Specifically, the evolution starts from uniform distribution, to truncated normal distribution with $\textrm{loc}=0.7$ and $\textrm{scale}=0.5$, to truncated normal distribution with $\textrm{loc}=0.7$ and $\textrm{scale}=0.25$, and eventually, to Dirac-delta distribution at $\alpha=0.7$. Legends $'\textrm{scale}=0.5'$ and $'\textrm{scale}=0.25'$ are used to differentiate the two truncated normal distributions. Detailed information on these distributions and meaning of parameters can be found in the Python open source package \texttt{scipy.stats}~\cite{virtanen2020scipy}.}
    \label{fig: alpha distribution}
\end{figure}

\begin{figure}[ht!]
    \centering
    \includegraphics[width=1\linewidth]{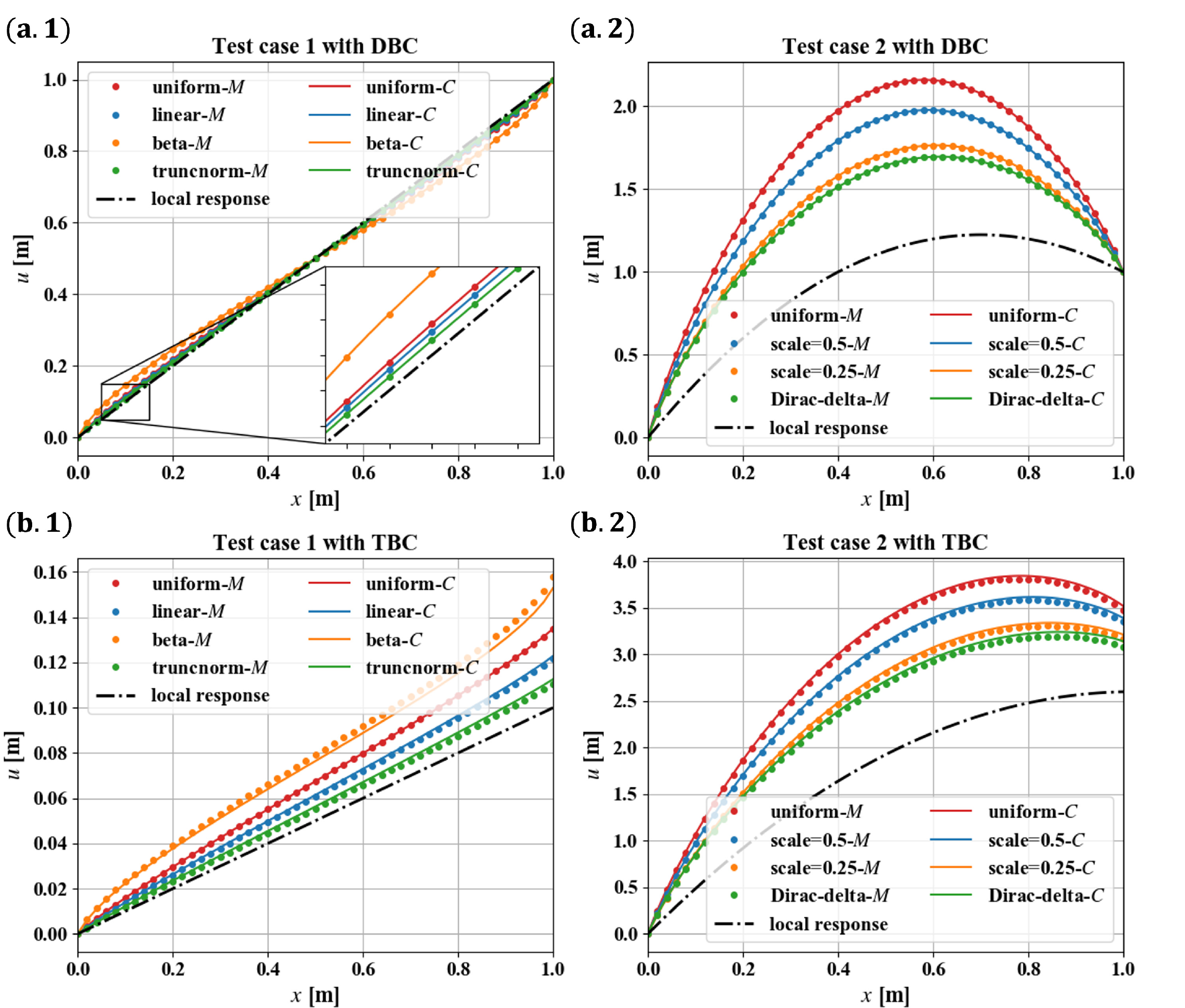}
    \caption{Numerical results of the predicted displacement distribution. 
    (a.1) and (a.2) show results for test case 1 and 2 under DBC; (b.1) and (b.2) show results for test case 1 and 2 under TBC. Results labeled $'-M'$ refer to predictions obtained using the MSLM model, while the label $'-C'$ indicates results provided by the continuous theory of DO-NET. The black dashed-dotted line indicates results obtained from a the local 1D system.}
    \label{fig: displacement_results}
\end{figure}

\begin{figure}[ht!]
    \centering
    \includegraphics[width=1\linewidth]{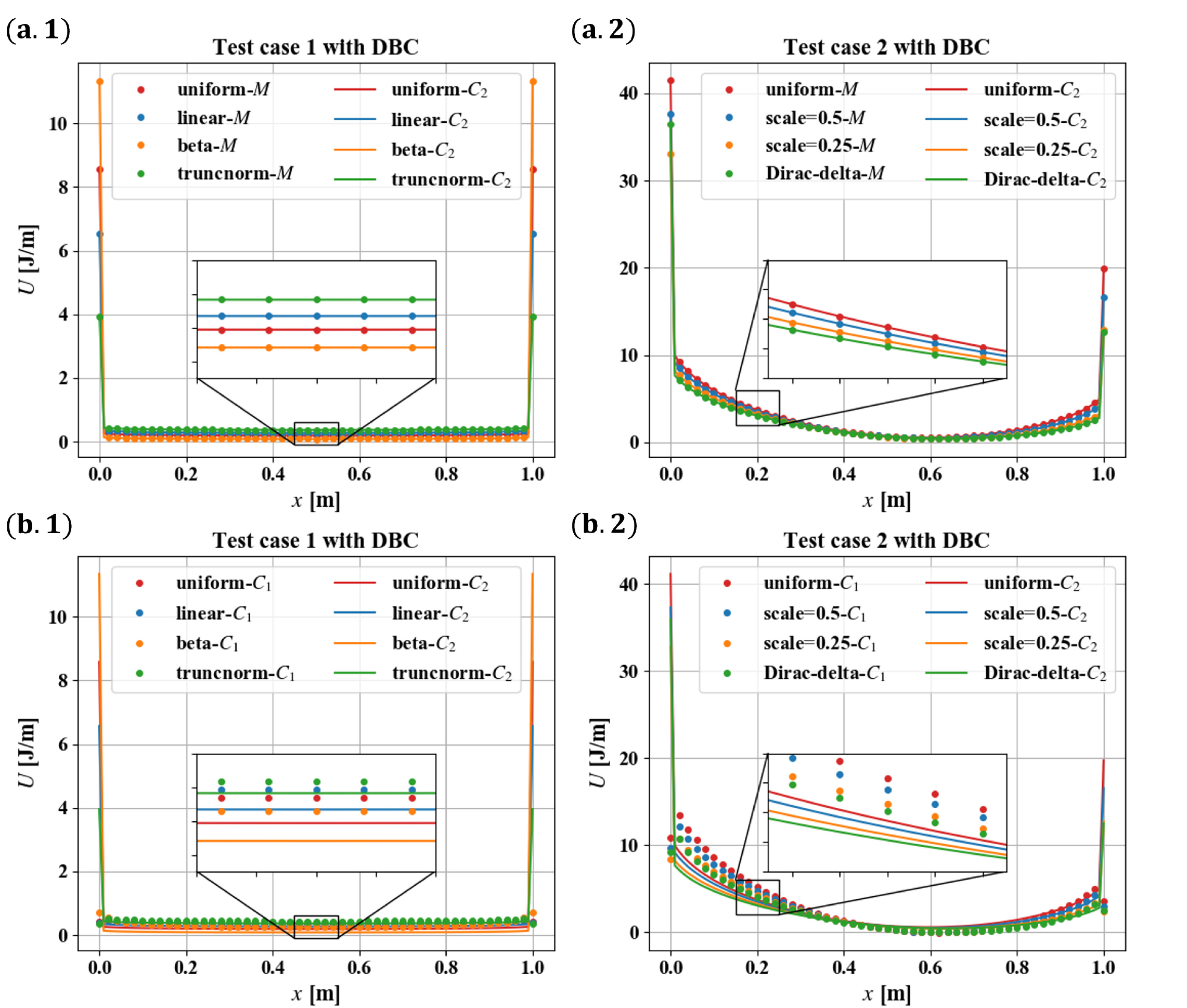}
    \caption{Potential energy density under DBC. (a.1) and (a.2) show simulation results of test case 1 and 2, respectively; (b.1) and (b.2) show simulation results of test case 1 and 2, respectively. (a.1) and (a.2) compare simulation results between $\mathbb{U}_i^M$ and $\mathbb{U}^{C_2}$ while (b.1) and (b.2) compare results between $\mathbb{U}^{C_1}$ and $\mathbb{U}^{C_2}$. Simulation results based on $\mathbb{U}^{M}$, $\mathbb{U}^{C_1}$, and $\mathbb{U}^{C_2}$ are labeled with $'M'$, $'{C_1}'$, and $'{C_2}'$, respectively. (a.1) and (a.2) show that $\mathbb{U}^{M}$ and $\mathbb{U}^{C_2}$ are in excellent agreement, while (b.1) and (b.2) show that $\mathbb{U}^{C_1}$ and $\mathbb{U}^{C_2}$ are not (due to the extra boundary energy terms $\mathbb{U}^{b}$ defined in $\mathbb{U}^{C_2}$).}
    \label{fig: PED_DBC}
\end{figure}

\begin{figure}[ht!]
    \centering
    \includegraphics[width=1\linewidth]{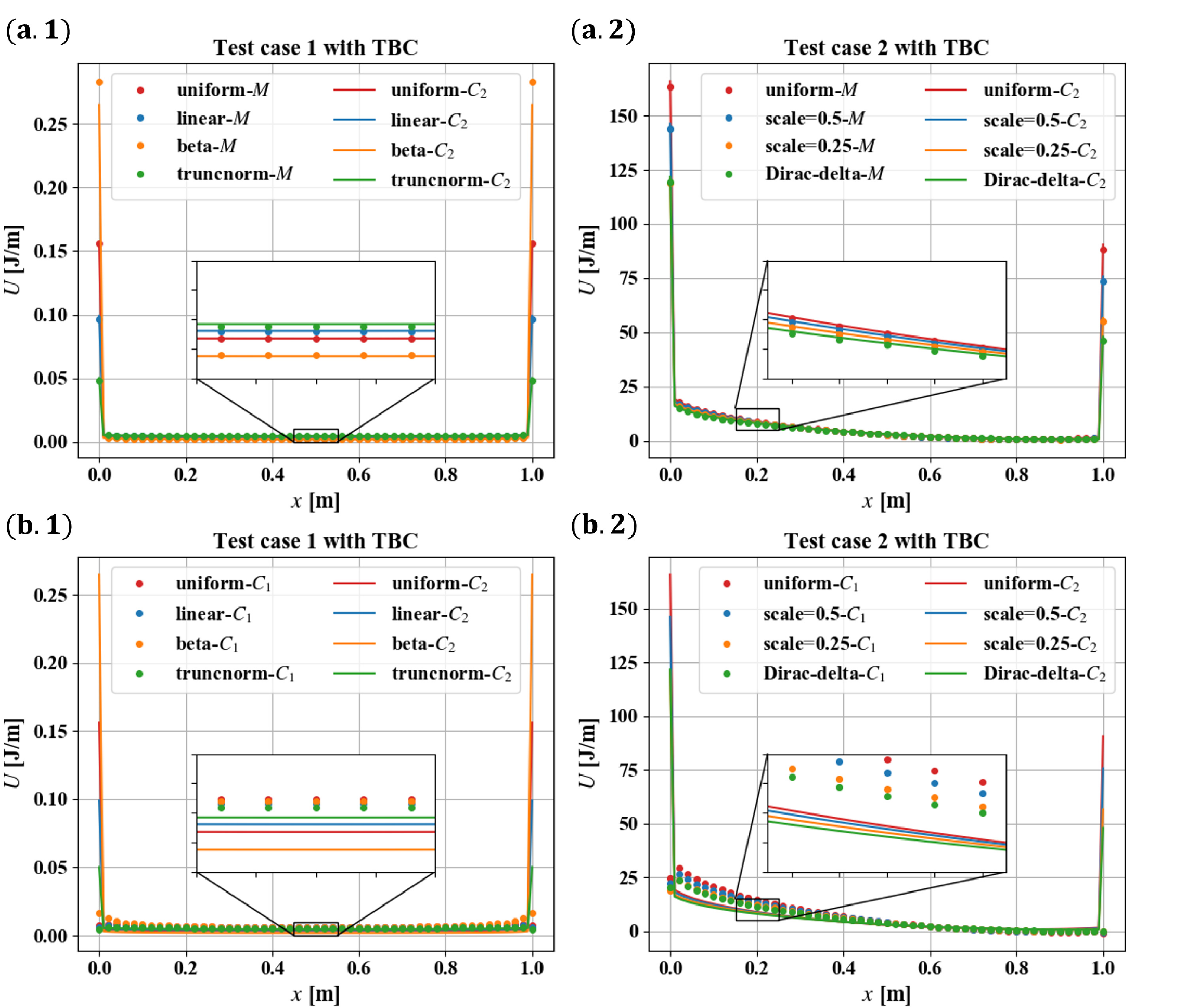}
    \caption{Potential energy density simulation results under TBC. (a.1) and (a.2) show simulation results of test case 1 and 2, respectively; (b.1) and (b.2) show simulation results of test case 1 and 2 respectively. In particular, (a.1) and (a.2) compare simulation results between $\mathbb{U}_i^M$ and $\mathbb{U}^{C_2}$; (b.1) and (b.2) compare simulation results between $\mathbb{U}^{C_1}$ and $\mathbb{U}^{C_2}$. Simulation results based on $\mathbb{U}^{M}$, $\mathbb{U}^{C_1}$, and $\mathbb{U}^{C_2}$ are labeled with $'M'$, $'{C_1}'$, and $'{C_2}'$, respectively. (a.1) and (a.2) show that $\mathbb{U}^{M}$ and $\mathbb{U}^{C_2}$ are in excellent agreement, while (b.1) and (b.2) show that $\mathbb{U}^{C_1}$ and $\mathbb{U}^{C_2}$ are not (due to the extra boundary energy terms $\mathbb{U}^{b}$ defined in $\mathbb{U}^{C_2}$).}
    \label{fig: PED_TBC}
\end{figure}

In this section, we present the response of the DO-NET and MSLM for different loading conditions and order distributions. More specifically, we consider the following different test cases:
\begin{itemize}
    \item \textbf{Test case 1:} four different continuous distributions of the order $\alpha$ with support in the closed interval $[0,1]$ are evaluated. Specifically, we consider the uniform, linear, beta, and truncated normal distributions. Their respective plot is provided in Fig.~(\ref{fig: alpha distribution}-a). The purpose of this test case is to explore the impact of the different $\alpha-$distributions on the nonlocal response and the ability of the DO-NET to capture different distributions. In all cases, the external load is applied at the end-point via a DBC and a TBC, which are given as $u(L)=1\textrm{m}$ and $T(L)=10\textrm{N}$.
    
    \item \textbf{Test case 2:} different $\alpha-$distributions are evaluated (see Fig.~(\ref{fig: alpha distribution}-b)). The purpose of this test case is to analyze the reduction of the DO model to a CO model via transition of the $\alpha-$distribution from a uniform distribution to a dirac-delta distribution centered at $\alpha_0$, that is, $\delta(\alpha-\alpha_0)$. The evolution of the uniform distribution to $\delta(\alpha-\alpha_0)$ is simulated through a series of normal distributions centered at $\alpha_0$ with reducing scale. 
    In this test case, a uniformly distributed axial force $F(x)=5\textrm{N}$ is applied on the body along with the DBC and TBC used in test case 1, that is, $u(L)=1\textrm{m}$ and $T(L)=10\textrm{N}$, respectively.
\end{itemize}
The properties of the different distributions are provided in Table~\ref{tab: 1}. Note that while the above distributions were chosen to validate the proposed MSLM and DO-NET under diverse conditions, most of these distributions have direct physical interpretation in real-world applications such as, for example, heterogeneous impurity distribution in graded junctions~\cite{kennedy1968measurement}, functionally graded materials with linearly varying properties~\cite{jain2004crack}, and even alloys with log-normal distributed grain size~\cite{bai2019grain}. By exploring eight different distributions of $\alpha$, we intend to provide a variety of conditions that could serve as fertile ground to identify possible applications across different fields involving multiscale nonlocal problems. The externally applied loads and boundary conditions for both the cases are schematically illustrated in Fig.~(\ref{fig: 1D rod}). The left-boundary of the rod is fixed, that is, $u(0)=0$. As for the numerical discretization, $n_\alpha=100$ points for the DO interval $\alpha\in[0,1]$ and $n=100$ points for the spatial domain of the 1D rod $[0,1]$m were adopted for the discretization. Both cases used uniformly distributed stencils. The numerical results are presented in Figures~(\ref{fig: displacement_results})-(\ref{fig: PED_TBC}) in terms of the displacement response and potential energy densities.

\begin{table}[!htbp]
    \centering
    \begin{tabular}{||c|c|
    >{\centering\arraybackslash}p{2cm}|
    >{\centering\arraybackslash}p{2cm}|
    >{\centering\arraybackslash}p{2cm}|
    >{\centering\arraybackslash}p{2.1cm}||}
    \hline
        \multirow{4}{6em}{\textbf{Test case 1}} & \textbf{\textit{Properties}} & \textbf{\textit{Uniform}} & \textbf{\textit{Linear}} & \textbf{\textit{Beta}} & \textbf{\textit{Truncnorm}} \\\cline{2-6}
        & \textbf{\textit{Mean}} & $0.5000$ & $0.6667$ & $0.2857$ & $0.8359$ \\\cline{2-6}
        & \textbf{\textit{Median}} & $0.5000$ & $0.7071$ & $0.2644$ & $0.8517$ \\\cline{2-6}
        & \textbf{\textit{Mode}} & - & $1$ & $0.2$ & $0.9$ \\\cline{2-6}
        & \textbf{\textit{Standard deviation}} & $0.2887$ & $0.2357$ & $0.1597$ & $0.1095$ \\
    \hline
        \multirow{4}{6em}{\textbf{Test case 2}} & \textit{\textbf{Properties}} & \textit{\textbf{Uniform}} & \textit{\textbf{Scale=0.5}} & \textit{\textbf{Scale=0.25}} & \textit{\textbf{Dirac-delta}} \\\cline{2-6}
        & \textit{\textbf{Mean}} & $0.5000$ & $0.5578$ & $0.6472$ & $0.7$ \\\cline{2-6}
        & \textit{\textbf{Median}} & $0.5000$ & $0.5775$ & $0.6646$ & $0.7$ \\\cline{2-6}
        & \textit{\textbf{Mode}} & - & $0.7$ & $0.7$ & $0.7$ \\\cline{2-6}
        & \textit{\textbf{Standard deviation}} & $0.2887$ & $0.2665$ & $0.2041$ & $0$ \\
    \hline
    \end{tabular}
    \caption{Measures of central tendency corresponding to the different $\kappa(\alpha)$ in Fig.~(\ref{fig: alpha distribution}).}
    \label{tab: 1}
\end{table}

Fig.~(\ref{fig: displacement_results}) shows the displacement fields of the DO-NET and MSLM obtained for all test cases. Fig.~(\ref{fig: displacement_results}a.1) and Fig.~(\ref{fig: displacement_results}a.2) present the results obtained for the DBC in test cases 1 and 2, respectively. Fig.~(\ref{fig: displacement_results}b.1) and Fig.~(\ref{fig: displacement_results}b.2) present the results obtained for the TBC in test cases 1 and 2, respectively. A detailed analysis of these results leads to the following observations and conclusions:
\begin{itemize}
    \item As evident from Fig.~(\ref{fig: displacement_results}a.1) and Fig.~(\ref{fig: displacement_results}a.2), the displacement fields simulated via the DO-NET and MSLM, for both the test cases 1 and 2 under DBC, are in excellent agreement with each other. Fig.~(\ref{fig: displacement_results}b.1) and Fig.~(\ref{fig: displacement_results}b.2) show that there is a small difference between the response of the DO-NET and MSLM for the TBC. 
    Nonetheless, the maximum point-wise difference between the DO-NET and MSLM response is less than $2\%$ (of either the DO-NET or MSLM response), for all the $\alpha-$distributions, which indicates a good match. We merely note that the difference between the two formulations under TBC is larger than the difference under DBC because the whole derivation of MSLM and its equivalence with TBC in DO-NET are based on the rectangular approximation (in \S\ref{sssec: 4.1.1}). While this approximation brings error at the boundaries when applying TBC, the DBC case does not involve approximation of the forces at the boundary points and thus accumulates smaller errors. The close match between the response of the DO-NET and MSLM for all the test cases demonstrates the equivalence of their EOMs in Eq.(S36,\ref{eq: 1D EOM MSLM continuum limit}) and validates the MSLM.
    
    \item The results in Fig.~(\ref{fig: displacement_results}a.1) and Fig.~(\ref{fig: displacement_results}b.1), which correspond to the test case 1, suggest that an increase in the degree of nonlocality leads to a greater distortion of the displacement field of the nonlocal solid with respect to the local solid (obtained for the distribution $\kappa(\alpha)=\delta(\alpha-1)$), under the same loads and boundary conditions. This is a direct result of the softening effect of the solid when subject to an increasing degree of nonlocality \cite{carpinteri2014nonlocal,patnaik2020generalized,patnaik2020ritz}.
    
    In order to better understand the above conclusions, we first discuss a possible approach to compare the degree of nonlocality of different $\alpha-$distributions. Recall that all the $\kappa(\alpha)$ are normalized, that is, $\int_0^1 \kappa(\alpha) \mathrm{d}\alpha = 1$; in other terms, the area under all $\kappa(\alpha)$ curves in Fig.~(\ref{fig: alpha distribution}) is equal to 1. The measures of central tendency of the different $\kappa(\alpha)$, provided in Table~\ref{tab: 1}, suggest that the \texttt{beta} distribution predominantly carries information from lower values of $\alpha$, followed by the \texttt{uniform} and \texttt{linear} distributions which carry information from progressively higher values of $\alpha$. The \texttt{truncnorm} distribution derives the maximum information from the highest values of $\alpha$. This trend is reflected from both the mean ($\alpha_\mu$) and median ($\alpha_{0.5}$) of the distributions which, starting from the minimum values seen in the \texttt{beta} distribution, increase progressively for the \texttt{uniform}, \texttt{linear}, and \texttt{truncnorm} distributions. Recall also that a lower value of the fractional-order in constant fractional-order nonlocal theories is indicative of a higher degree of nonlocality \cite{patnaik2020generalized,patnaik2020ritz}. It immediately follows that, the degree of nonlocality increases with decreasing $\alpha_\mu$ and $\alpha_{0.5}$.
    
    The results in Fig.~(\ref{fig: displacement_results}a.1) and Fig.~(\ref{fig: displacement_results}b.1) are consistent with the above discussion. As evident from these results, the \texttt{beta} distribution (which has the lowest $\alpha_\mu$ and $\alpha_{0.5}$) leads to the most pronounced softening effect, while the \texttt{truncnorm} distribution (which has the highest $\alpha_\mu$ and $\alpha_{0.5}$) leads to the lowest effect in terms of softening.
    
    \item Following the above discussion, we expect that in test case 2, the \texttt{uniform} distribution will be associated with the strongest softening effect, followed by the \texttt{truncnorm} distribution with scale $0.5$ and the \texttt{truncnorm} distribution with scale $0.25$. Finally, the CO distribution $\kappa(\alpha)=\delta(\alpha-0.7)$ is expected to be affected the least from softening effects (compared with the local response). Indeed, the results in Fig.~(\ref{fig: displacement_results}a.2) and Fig.~(\ref{fig: displacement_results}b.2) are consistent with the above discussion. Also of interest is that, for this test case, the response obtained by the uniform $\kappa(\alpha)$ appears to converge to the CO response, when $\kappa(\alpha)$ evolves from the uniform distribution to the Dirac-delta distribution via the sequence of truncated normal distributions.
    
    \item The above two points suggest that the consistent softening response, with increasing degree of nonlocality, is observed independently of the loading and boundary conditions. This observation is in contrast with strain-driven approaches which are typically ill-posed for different loading and boundary conditions and can lead to inconsistent predictions \cite{challamel2014nonconservativeness}.
\end{itemize}

Fig.~(\ref{fig: PED_DBC}) and Fig.~(\ref{fig: PED_TBC}) present the potential energy densities obtained via the DBC and TBC, respectively. In each case, we compute the potential energy density using three previously derived expressions: $\mathbb{U}^{C_1}$ and $\mathbb{U}^{C_2}$ defined at continuum level, and $\mathbb{U}_i^{M}$ defined at discrete level. Accordingly, we use the displacement fields obtained via the DO-NET to compute $\mathbb{U}^{C_1}$ and $\mathbb{U}^{C_2}$, and the displacement field obtained via MSLM to compute $\mathbb{U}_i^{M}$. In sub-figures (a.1) and (a.2), we compare the potential energy density obtained via the MSLM ($\mathbb{U}_i^{M}$) and the continuum expression $\mathbb{U}^{C_1}$ for the two test cases 1 and 2, respectively. Next, in the sub-figures (b.1) and (b.2), we compare the two continuum expressions $\mathbb{U}^{C_1}$ and $\mathbb{U}^{C_2}$ for the two test cases 1 and 2, respectively. The results presented in Fig.~(\ref{fig: PED_DBC}) and Fig.~(\ref{fig: PED_TBC}) lead to the following observations and remarks:
\begin{itemize}
    \item From Fig.~(\ref{fig: PED_DBC})(a.1),(a.2) and Fig.~(\ref{fig: PED_TBC})(a.1),(a.2), the match between $\mathbb{U}^{C_2}$ and $\mathbb{U}_i^{M}$ is excellent for all the points within the domain of the rod, and for the boundaries located at $x\in\{0,L\}$. This is not surprising since the displacement fields computed by DO-NET and MSLM are in excellent agreement and $\mathbb{U}^{C_2}$ was obtained via the continuum limit of $\mathbb{U}_i^{M}$ (see Eqs.~(\ref{eq: continuum limit U_i^M},\ref{eq: U^C2})). Note that, at the boundary points, we compute $\mathbb{U}^{C_2}(0)$ and $\mathbb{U}^{C_2}(L)$ by also considering the boundary energy terms $\mathbb{U}^{b}(0)$ and  $-\mathbb{U}^{b}(L)$, respectively (see Eq.~(\ref{eq: Pi^M and Pi^C1})). Recall that boundary energy contributions were isolated in \S\ref{sssec: 4.1.2} by ensuring a consistency between the numerical discretization adopted for the continuum model and the MSLM. By considering the boundary energy terms $\mathbb{U}^b$, we observe a concentration of the deformation energy at the boundaries. This behaviour is also present in the lattice model ($\mathbb{U}_i^{M}$). It appears that the strength of this energy concentration is proportional to the degree of nonlocality, that is, it is maximum for the $\kappa(\alpha)$ with the maximum degree of nonlocality (\texttt{beta} distribution in test case 1 and \texttt{uniform} distribution in test case 2). Note that the boundary energy terms in Eq.~(\ref{eq: Pi^M and Pi^C1}) were related to the so-called surface energy existing in micro/nano structures. At this scale, the impact of the long-range interactions (typically resulting from atomic interactions) operating at the surface is significant, since the surface thickness is comparable to the length-scale of the system~\cite{li2020contribution}.

    \item Unlike the above observation, the potential energy densities $\mathbb{U}^{C_1}$ and $\mathbb{U}^{C_2}$ presented in Fig.~(\ref{fig: PED_DBC})(b.1),(b.2) and Fig.~(\ref{fig: PED_TBC})(b.1),(b.2) present a poor match. While, at a first glance, this difference might seem to suggest an inconsistency between the two continuum expressions $\mathbb{U}^{C_1}$ and $\mathbb{U}^{C_2}$, in practice it should be en expected outcome. Indeed, this difference is a direct result of the definitions for $\mathbb{U}^{C_1}$ and $\mathbb{U}^{C_2}$ that were aimed at capturing different underlying phenomena. More specifically, while $\mathbb{U}^{C_1}$ was defined at a strictly continuum level (via Eq.~(\ref{eq: Pi^C1})), $\mathbb{U}^{C_2}$ was defined specifically (by leveraging the MSLM and the numerical technique in \S\ref{sssec: 4.1.2}) to capture the surface energy densities that are typically observed at very fine (e.g. atomic) scales. In the following point, we provide some additional observations that clarify this subtle yet important difference. 
    
    \item First, observe from the different nonlocal response in Fig.~(\ref{fig: displacement_results}) that the displacement of the system, obtained via both the DO-NET and MSLM, is at least $C^2$ continuous (apparent from the lack of inflection points). Hence, it immediately follows that the strain and stress definitions at continuum level, and consequently the potential energy density ($\mathbb{U}^{C_1}$) are smooth (also evident from sub-figures (b.1) and (b.2) in Figs.~(\ref{fig: PED_DBC}, \ref{fig: PED_TBC})). Next, note that although $\mathbb{U}^{C_1}$ and $\mathbb{U}^{C_2}$ differ at a local (point-wise) level, at a global level the total potential energies $\Pi^{C_1}$, $\Pi^{C_2}$, and $\Pi^{M}$ should be equivalent (according to Eq.~(\ref{eq: Pi^M and Pi^C1})). This can be seen in Table~\ref{tab: 2} and Table~\ref{tab: 3} where the total potential energies computed for the different test cases are provided. As evident from these results, the total potential energy obtained from the three different potential energy densities, match very well and are within a 1\% difference from each others. It can be envisioned that, since $\mathbb{U}^{C_2}$ isolates the surface energy contributions from the total potential density and the total potential energies are the same irrespective of the specific definitions (as it should be from a physical perspective), the remaining energy is \textit{redistributed} in the 1D system when using the expression for $\mathbb{U}^{C_2}$. This is also evident from the insets within the sub-figures (b.1) and (b.2), where one can observe that $\mathbb{U}^{C_1}>\mathbb{U}^{C_2}$ in the selected region of the 1D nonlocal rod.
    
    \begin{table}[!htbp]
    \centering
    \begin{tabular}{||c|c|
    >{\centering\arraybackslash}p{1.8cm}|
    >{\centering\arraybackslash}p{1.8cm}|
    >{\centering\arraybackslash}p{1.9cm}|
    >{\centering\arraybackslash}p{2.1cm}||}
    \hline
        \multirow{4}{5.5em}{\textbf{Test case 1}} & \textbf{\textit{Total Potential Energy}} & \textbf{\textit{Uniform}} & \textbf{\textit{Linear}} & \textbf{\textit{Beta}} & \textbf{\textit{Truncnorm}}  \\\cline{2-6}
        & $\Pi^{C_1}$ & $0.3630$ & $0.4029$ & $0.3087$ & $0.4435$ \\\cline{2-6}
        & $\Pi^{C_2}$ & $0.3632$ & $0.4057$ & $0.3139$ & $0.4435$ \\\cline{2-6}
        & $\Pi^{M}$ & $0.3702$ & $0.4110$ & $0.3161$ & $0.4521$ \\
    \hline
        \multirow{4}{5.5em}{\textbf{Test case 2}} & \textbf{\textit{Total Potential Energy}} & \textit{\textbf{Uniform}} & \textit{\textbf{Scale=0.5}} & \textit{\textbf{Scale=0.25}} & \textit{\textbf{Dirac-delta}} \\\cline{2-6}
        & $\Pi^{C_1}$ & $3.0430$ & $2.7478$ & $2.4053$ & $2.3062$ \\\cline{2-6}
        & $\Pi^{C_2}$ & $3.0650$ & $2.7675$ & $2.4229$ & $2.3303$ \\\cline{2-6}
        & $\Pi^{M}$ & $3.1637$ & $2.8531$ & $2.4934$ & $2.3981$ \\
    \hline
    \end{tabular}
    \caption{Total potential energy computed in the case of displacement boundary condition (DBC). $C_1$, $C_2$, and $M$ represent total potential energy formulation $\Pi^{C_1}$, $\Pi^{C_2}$, and $\Pi^{M}$, respectively.}
    \label{tab: 2}
\end{table}

\begin{table}[ht!]
    \centering
    \begin{tabular}{||c||c|
    >{\centering\arraybackslash}p{1.8cm}|
    >{\centering\arraybackslash}p{1.8cm}|
    >{\centering\arraybackslash}p{1.9cm}|
    >{\centering\arraybackslash}p{2.1cm}||}
    \hline
        \multirow{4}{5.5em}{\textbf{Test case 1}} & \textbf{\textit{Total Potential Energy}} & \textbf{\textit{Uniform}} & \textbf{\textit{Linear}} & \textbf{\textit{Beta}} & \textbf{\textit{Truncnorm}} \\\cline{2-6}
        & $\Pi^{C_1}$ & $0.6609{\times}10^{-3}$ & $0.6081{\times}10^{-3}$ & $0.7212{\times}10^{-3}$ & $0.5624{\times}10^{-3}$ \\\cline{2-6}
        & $\Pi^{C_2}$ & $0.6668{\times}10^{-3}$ & $0.6123{\times}10^{-3}$ & $0.7333{\times}10^{-3}$ & $0.5650{\times}10^{-3}$ \\\cline{2-6}
        & $\Pi^{M}$ & $0.6752{\times}10^{-3}$ & $0.6082{\times}10^{-3}$ & $0.7907{\times}10^{-3}$ & $0.5529{\times}10^{-3}$ \\
    \hline
        \multirow{4}{5.5em}{\textbf{Test case 2}} & \textbf{\textit{Total Potential Energy}} & \textit{\textbf{Uniform}} & \textit{\textbf{Scale=0.5}} & \textit{\textbf{Scale=0.25}} & \textit{\textbf{Dirac-delta}} \\\cline{2-6}
        & $\Pi^{C_1}$ & $7.2925$ & $6.8022$ & $6.1958$ & $6.0073$ \\\cline{2-6}
        & $\Pi^{C_2}$ & $7.3022$ & $6.8153$ & $6.2127$ & $6.0326$ \\\cline{2-6}
        & $\Pi^{M}$ & $7.2712$ & $6.7837$ & $6.1734$ & $5.9547$ \\
    \hline
    \end{tabular}
    \caption{Total potential energy computed under traction boundary condition. $C_1$, $C_2$, and $M$ represent total potential energy formulation $\Pi^{C_1}$, $\Pi^{C_2}$, and $\Pi^{M}$, respectively.}
    \label{tab: 3}
\end{table}

    \item To further clarify the latter comment on the energy redistribution obtained via $\mathbb{U}^{C_2}$, let us consider an alternative definition of $\mathbb{U}_{i}^{M}$ which also conserves the total deformation energy of the 1D MSLM:
    \begin{equation}\label{eq: U_i^M1} 
        \mathbb{U}_i^{M_1}=\frac{1}{4}\sum_{p=0}^{i}\sum_{q=i}^{n}\frac{M_1(i)}{q-p}\left(\int_{0}^{1}\kappa(\alpha)k_{pq}(\alpha)\textrm{d}\alpha\right)\left(u_p-u_q\right)^2
    \end{equation}
    where $M_1(i)=1$ for body points $0<i<n$ and $M_1(i)=1/2$ for two boundary points $i=0,n$. Recall that Eq.~(\ref{eq: U_i^M}) computes $\mathbb{U}_{i}^{M}$ at a point $i$ by taking half of the energy of all the springs with one end fixed at the point $i$ into account ($k_{ij}$ with $0 \leq j \leq n$). Contrary to Eq.~(\ref{eq: U_i^M}), Eq.~(\ref{eq: U_i^M1}) computes $\mathbb{U}_{i}^{M_1}$ by considering all springs that contain the point $i$ within their span ($k_{pq}$ with $0 \leq p \leq i$ and $i \leq q \leq n$). More specifically, under this definition, for a given spring connecting two point $p$ and $q$, the spring energy will be distributed not only at these two points, but also at all the points in-between them. Figure~(\ref{fig: U^M1 and U^C1}) shows the comparison between $\mathbb{U}_{i}^{M_1}$ and $\mathbb{U}^{C_1}$ under DBC. It can be seen that compared with $\mathbb{U}^{C_2}$ and $\mathbb{U}_{i}^{M}$ in Fig.~(\ref{fig: PED_DBC}) and Fig.~(\ref{fig: PED_TBC}), $\mathbb{U}_{i}^{M_1}$ does not possess any boundary energy concentration. Although this alternative definition does not exactly match $\mathbb{U}^{C_1}$, the scale and distribution of the potential energy density is much closer than the previous results in Fig.~(\ref{fig: PED_DBC}) and Fig.~(\ref{fig: PED_TBC}). Based on the above discussions, it is reasonable to attribute the difference between $\mathbb{U}^{C_1}$ and $\mathbb{U}^{C_2}$ to different definitions of the potential energy and not to any inconsistency between the DO-NET and the MSLM approaches.
    
    \begin{figure}[ht!]
    \centering
    \includegraphics[width=\linewidth]{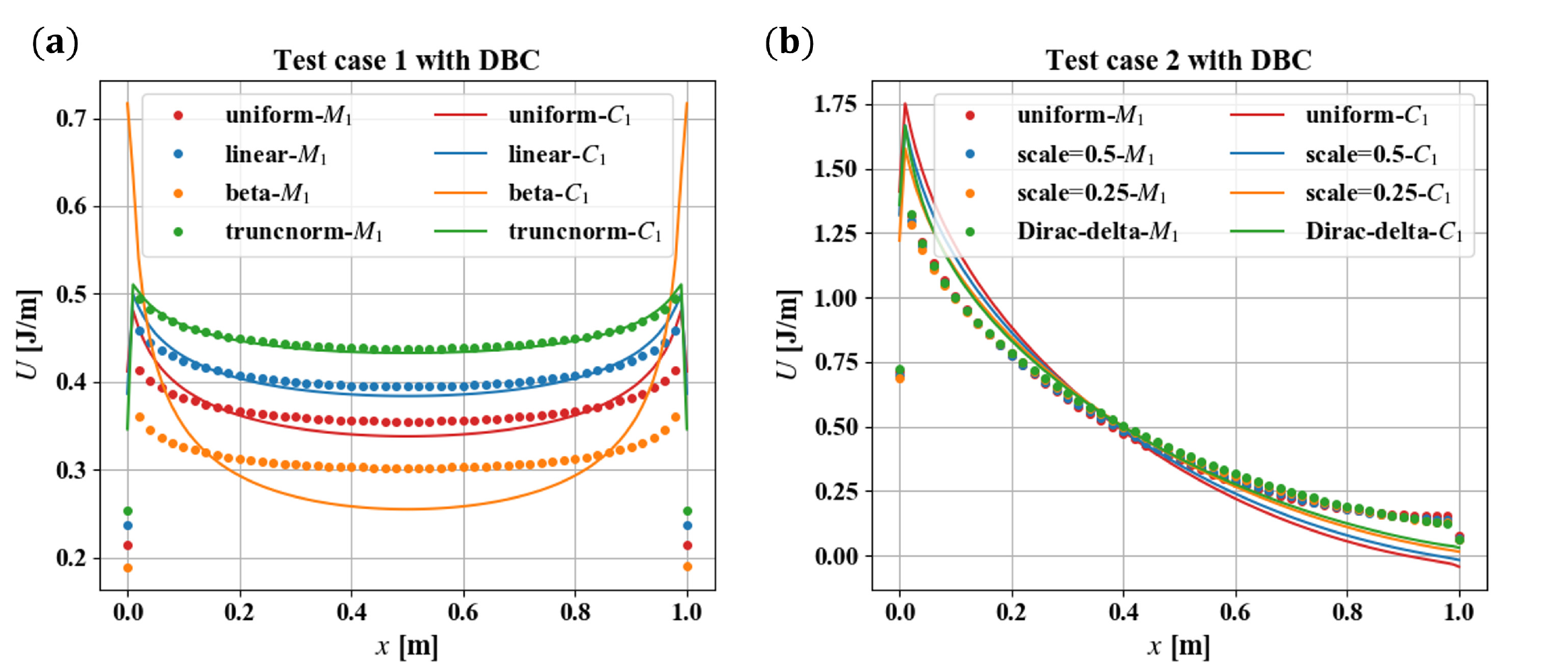}
    \caption{Numerical estimates of the potential energy density $\mathbb{U}_{i}^{M_1}$ and $\mathbb{U}^{C_1}$ under DBC. (a) shows results for test case 1, and (b) shows results for test case 2. The difference between $\mathbb{U}_{i}^{M_1}$ and $\mathbb{U}^{C_1}$ is found to be much smaller than the previous case in Fig.~(\ref{fig: PED_DBC}), especially at both boundary points. This indicates that an alternative definition ($\mathbb{U}_{i}^{M_1}$) can reduce the inconsistency of the potential energy density between MSLM and DO-NET.}
    \label{fig: U^M1 and U^C1}
    \end{figure}
\end{itemize}
  
\begin{figure}[ht!]
    \centering
    \includegraphics[width=1\linewidth]{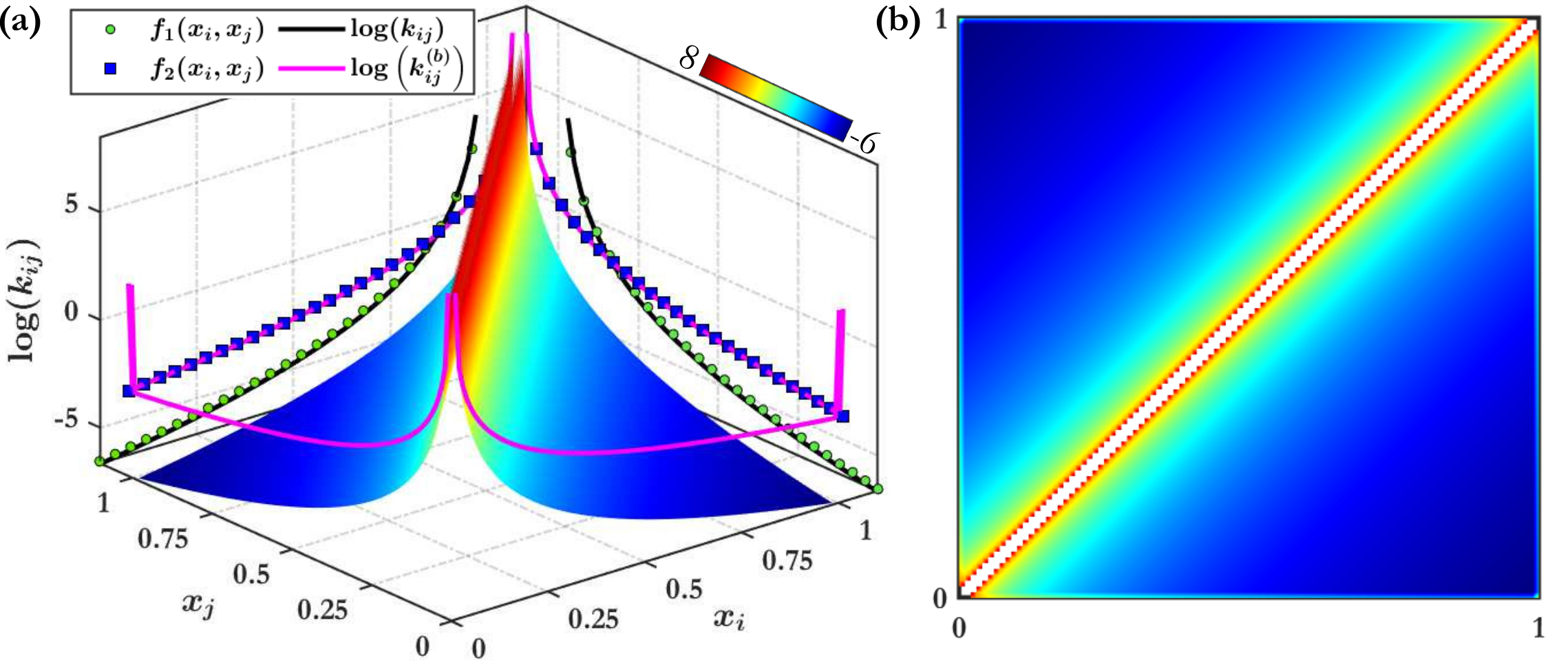}
    \caption{Elastic spring stiffness distribution. (a) shows scatter plot of $\textrm{log}(k_{ij})$. Green circle and blue square lines are the power-law decay function $f_1(x_i,x_j)$ and $f_2(x_i,x_j)$ defined in Eq.~(\ref{eq: f_1 and f_2}), respectively. Black and pink lines are projections of body spring stiffness $\mathrm{log}(k_{ij}^{\alpha})$ and boundary spring stiffness $\mathrm{log}(k_{ij}^{(b)})$; (b) shows overall distribution of $\textrm{log}(k_{ij}^{\alpha})$ using heat map plot.}
    \label{fig: k_ij}
\end{figure}

In addition to the above numerical results, we present yet another argument that support the equivalence between DO-NET and MSLM, namely, the distribution of the elastic spring stiffness $k_{ij}^{\alpha}$. Here, we present the distribution of $k_{ij}$ obtained from the truncated normal distribution in Fig.~(\ref{fig: alpha distribution}-a). Fig.~(\ref{fig: k_ij}) shows detailed results of $k_{ij}^{\alpha}$. To facilitate the presentation, we consider $\textrm{log}(k_{ij}^{\alpha})$ in the plots presented in Fig.~(\ref{fig: k_ij}).
As evident from Fig.~(\ref{fig: k_ij}), $k_{ij}^{\alpha}$ decays symmetrically about any point within the nonlocal solid. The symmetry of $k_{ij}^{\alpha}$ is more evident from the 2D (top-view) projection of the surface plot in Fig.~(\ref{fig: k_ij}b). Recalling that the spring stiffness is a direct indicator of the strength of nonlocality, the decay in the nonlocal spring stiffness in the MSLM is analogous to the characteristics of the attenuation kernel of the DO-NET or of other types of nonlocal elasticity theories~\cite{eringen1972nonlocal,carpinteri2014nonlocal}. Observe that the boundary nonlocal stiffness terms ($k_{0j}^{\alpha},k_{i0}^{\alpha},k_{nj}^{\alpha},k_{in}^{\alpha}$) decay at a slower rate and are much stiffer than the nonlocal springs within the MSLM. In fact, it is also evident from Eq.~(\ref{eq: k_ij}) that the body spring stiffness decay via a power-law with exponent $-(2+\alpha)$ while the boundary spring stiffness decay predominantly with an exponent $-(1+\alpha)$. In order to better present this phenomenon, in Fig.~(\ref{fig: k_ij}-a) we have compared selected combinations of these stiffness with the following functions:
\begin{equation}\label{eq: f_1 and f_2}
    f_1(x_i,x_j)=\text{log}\left(\int_{0}^{1}\frac{\kappa(\alpha)}{|x_i-x_j|^{2+\alpha}}\textrm{d}\alpha\right),\quad
    f_2(x_i,x_j)=\text{log}\left(\int_{0}^{1}\frac{\kappa(\alpha)}{|x_i-x_j|^{1+\alpha}}\textrm{d}\alpha\right)
\end{equation}
The choice of these functions is also motivated from their appearance within the DO-NET constitutive relations in Eq.~(S38,40), respectively. Hence, from a different perspective, this comparison will also demonstrate an equivalence between the degree of nonlocality obtained via the MSLM and DO-NET. In Fig.~(\ref{fig: k_ij}-a), the function $f_1(x_i,x_j)$ is presented for points $\{x_i,x_j\}$ such that $x_i+x_j=1$, and compared against the corresponding curves projected from the $k_{ij}^{\alpha}$ surface plot at two different planes where either $x_i$ or $x_j$ are constant. The condition $x_i+x_j=1$ suggests that we extract the stiffness from the diagonal along the surface plot of $k_{ij}^{\alpha}$. Further, we have also compared the function $f_2(x_i,x_j)$ with the boundary stiffness $\{k_{i0}^{\alpha},k_{nj}^{\alpha}\}$ (denoted as $k^{(b)}_{ij}$ collectively in the legend of the plot). As evident, the match between the different functions and the MSLM stiffness is excellent. More importantly, given the fact that boundary springs are stiffer than body springs, under the same deformation field, boundary springs can store higher potential energy than body springs. Remarkably, this observation also supports the previous discussion on the surface effects and energy concentration at the boundary.

In conclusion, our theoretical formulation and numerical results have shown that on the one hand MSLM provides a very effective route to a physical interpretation of DO-NET, and on the other hand DO-NET can be regarded as an effective multiscale homogenization method for complex nonlocal systems. Specifically, for a complex MSLM with co-existing nonlocality shown in Fig.~(\ref{fig: MSLM}), the traditional approach to study its mechanical response is to first 1) obtain equations of motion by Newton's second law, and then 2) to solve the equations~\cite{jensen2003phononic}. Although this approach is straightforward and easy to use, the computational cost increases drastically with the increase in problem complexity and spectrum of spatial scales involved. A typical example that directly exemplifies this problem is the class of molecular dynamics simulations. Although nowadays molecular dynamics can simulate up to one billion particles~\cite{shibuta2017heterogeneity}, the overall spatial scale is still restricted to the nano or, at most, the micro scales due to the computational cost. To address this problem, multiscale methods such as extended finite element method (XFEM) and bridge domain method (BDM)~\cite{talebi2015concurrent} are proposed. Similarly, for large scale particle and lattice systems (as the MSLM in this study), direct numerical simulation is computationally expensive and becomes rapidly unfeasible for large systems. The DO-NET provide an effective and computationally efficient approach to model the system at the macro scales while still retaining a significant amount of information from the smaller scales.

%%%%%%%%%%%%%%%%%%%%%%%%%%%%%%%%%%%%%%%%%%%%%%%%%%%%%%%%%%%%%%%%
\section{Conclusions}\label{sec: 6}
This study focused on the identification and characterisation of the complex physical phenomenon of multiscale nonlocal elasticity. This mechanism arises from the coexistence of multiple material scales in complex heterogeneous structures such as, for example, layered composites, functionally graded and porous materials. In these classes of material and structural systems, heterogeneity localized at different scales produces nonlocal effects (of varying strength) subject to cross-interaction due to the overall multiscale nature of the problem. The identification of multiscale nonlocal elasticity, as a physical phenomenon resulting from the coexistence and interaction of either multiscale or nonlocal effects was shown to be well-captured by distributed-order operators. Indeed, by using a sample structural problem consisting in a layered multiscale nonlocal material, this study demonstrated that the DO operators can successfully capture multiscale effects, nonlocal effects, as well as the interaction between them. This latter observation motivated the development of a 3D distributed-order nonlocal elasticity theory (DO-NET) capable of modeling this type of generalized multiscale nonlocal continua. The DO-NET was derived from a nonlocal thermodynamic formulation where multiscale effects were modeled using DO derivatives. Important aspects of the analysis including the derivation of governing equations via variational principles, the assessment of the well-posed nature of the governing equations, and their ability to capture heterogeneity and anisotropy (contrary to classical nonlocal approaches developed in literature) were presented in detail. 

To further understand the ability of the DO-NET to capture the underlying multiscale nonlocal mechanisms as well as its effectiveness in modeling complex continua, we presented a generalized mass spring lattice model (MSLM) approach representing an equivalent system made of parallel distribution of long-range elastic springs. The stiffness of each long-range spring was obtained by discretizing the distributed order derivatives in the DO-NET governing equations and by leveraging the mathematical equivalence between the fractional-order Caputo and the fractional-order Marchaud derivative. By taking the continuum limit, the MSLM was proved to be equivalent to the proposed DO-NET in terms of linear momentum, potential energy, and traction boundary conditions. These theoretical analyses provided critical insights on the ability of the DO-NET to achieve consistent predictions free from paradoxical behavior or boundary effects (often found in classical nonlocal approaches), as well as the ability to capture surface effects and energy concentration (typical of multiscale effects). To further illustrate the equivalence between the DO-NET and MSLM approaches, and to present different phenomena typical of multiscale systems such as material softening, surface effects, and energy concentration, comprehensive numerical simulations were conducted. Numerical results have shown an excellent match between the two modeling approaches in terms of both the displacement field and the total potential energy. 
The equivalence between the DO-NET and the MSLM not only contributes to the understanding of the nonlocal mechanisms captured by the DO-NET but, more importantly, illustrates the outstanding potential of DO-NET to accurately model at continuum level complex multiscale nonlocal systems.

In conclusion, the physically-consistent and well-posed DO-NET approach to multiscale nonlocal continua provides a critical step to establish accurate and efficient fractional-order continuum mechanics approaches to modeling the response of real-world complex structures.\\

%%%%%%%%%%%%%%%%%%%%%%%%%%%%%%%%%%%%%%%%%%%%%%%%%%%%%%%%%%%%%%%%
\noindent \textbf{Acknowledgements:} The authors gratefully acknowledge the financial support of the National Science Foundation (NSF) under grants MOMS \#1761423 and DCSD \#1825837, and the Defense Advanced Research Project Agency (DARPA) under grant \#D19AP00052. S.P. acknowledges the financial support of the School of Mechanical Engineering, Purdue University, through the Hugh W. and Edna M. Donnan Fellowship. Any opinions, findings, and conclusions or recommendations expressed in this material are those of the author(s) and do not necessarily reflect the views of the National Science Foundation. The content and information presented in this manuscript do not necessarily reflect the position or the policy of the government. The material is approved for public release; distribution is unlimited.\\

%\section*{Competing Interests} 
\noindent \textbf{Competing interests:} The authors declare no competing interest.
%%%%%%%%%%%%%%%%%%%%%%%%%%%%%%%%%%%%%%%%%%%%%%%%%%%%%%%%%%%%%%%%

\bibliographystyle{naturemag}
\bibliography{ref}

\end{document}